\documentclass[preprint2]{aastex}

\shorttitle{\textit{AKARI} Observations of Brown Dwarfs I.}
\shortauthors{Yamamura et al.}

\begin{document}


\title{\textit{AKARI} Observations of Brown Dwarfs I.: \\
CO and CO$_2$ Bands in the Near-Infrared Spectra}


\author{Issei Yamamura}
\affil{Institute of Space and Astronautical Science (ISAS), JAXA,
Yoshino-dai 3-1-1, Chuo-ku, Sagamihara, Kanagawa 252-5210, Japan}
\email{yamamura@ir.isas.jaxa.jp}

\and

\author{Takashi Tsuji and Toshihiko Tanab\'e}
\affil{Institute of Astronomy, School of Science, The University of Tokyo,
2-21-1, Osawa, Mitaka, Tokyo, 181-0015, Japan}
\email{ttsuji@ioa.s.u-tokyo.ac.jp, ttanabe@ioa.s.u-tokyo.ac.jp}

%




\begin{abstract}
Near-infrared medium-resolution spectra of seven bright brown dwarfs
are presented. The spectra were obtained with the Infrared Camera (IRC)
on board the infrared astronomical satellite \textit{AKARI},
covering 2.5--5.0~$\mu$m with a
spectral resolution of approximately 120.
The spectral types of the objects range from L5 to T8, and enable us
to study the spectral evolution of brown dwarfs.
The observed spectra are in general consistent with the predictions
from the previous observations and photospheric models;
spectra of L-type dwarfs are characterized by continuum opacity from
dust clouds in the photosphere, while very strong molecular absorption bands
dominate the spectra in the T-type dwarfs. We find that the CO fundamental
band around 4.6~$\mu$m is clearly seen even in the T8 dwarf 2MASS J041519$-$0935,
confirming the presence of non-equilibrium chemical state in the atmosphere.
We also identify the CO$_2$ fundamental stretching-mode band at 4.2~$\mu$m
for the first time in the spectra of late-L and T-type brown dwarfs.

As a preliminary step towards interpretation of the data obtained 
by \textit{AKARI}, 
we analyze the observed spectra by comparing with the predicted ones
based on the Unified Cloudy Model (UCM). Although overall spectral
energy distributions (SEDs)
can be reasonably fitted with the UCM, observed CO and CO$_2$ bands 
in late-L and T-dwarfs are unexpectedly stronger than the 
model predictions assuming local thermodynamical equilibrium (LTE).
We examine the vertical mixing model and find that this model explains
the CO band at least partly in the T-dwarfs 2MASS J041519$-$0935 and
2MASS J055919$-$1404. The CO fundamental band also 
shows excess absorption against the predicted one in the L9 dwarf 
SDSS J083008$+$4828. Since CO is already
highly abundant in the upper photospheres of late-L dwarfs,
the extra CO by vertical mixing has little effect on the CO band
strengths, and the vertical mixing model cannot be applied to this L-dwarf. 
A more serious problem is that the significant enhancement of the CO$_2$ 
4.2~$\mu$m band in both the late-L and T dwarfs cannot be explained at all 
by the vertical mixing model.
The enhancement of the CO$_2$ band remains puzzling.

\end{abstract}



\section{Introduction}

 Brown dwarfs are defined as the objects that were born as
isolated objects but were not massive enough to ignite hydrogen nuclear
burning \citep{Burrows01}. They are particularly of interest
for their extremely low-temperature atmospheres in relation
to giant planets.
Observational studies of brown dwarfs have been extensively
carried out since the first definitive identification of 
a genuine brown dwarf Gl~229B (of the class now known as T-dwarf)
by \citet{Nakajima95}. The first L-type
brown dwarf GD 165B was reported by \citet{Becklin88} although
it was debated for some time whether it was a star or a brown dwarf.
This very red object is important in that the presence of dust
layer in the photosphere was recognized  for the first time by this
object \citep{Tsuji96}.  Spectroscopic observations in
the infrared regime are the most essential tools to obtain
physical and chemical information of brown dwarfs, through
absorption bands of various molecules in the atmospheres. 

Carbon monoxide (CO) plays a decisive role in
determining the major characteristics of cool stars because of
its very large dissociation energy \citep{Russell34}.
In brown dwarfs, the CO first overtone band at 2.3 $\mu$m
is observed until early-T type sources (see e.g.,
\citet{Geballe02} and \citet{Burgasser06b}).
The role of CO, however, changes in very cool dwarfs:
carbon resides mostly in CH$_4$ rather than in CO in a very 
cool (e.g. $T \approx 1000$K) and high density 
(e.g. $\log P_\mathrm{g} \approx 6.0$) environment 
\citep{Tsuji64}, and CO no longer plays any critical 
role under such a circumstance. Although no celestial object outside
our Solar system having 
such a physical condition had been known, it was 
found at last that such a case is actually realized in the brown 
dwarf Gl~229B showing strong CH$_4$ bands \citep{Oppenheimer95}.

However, nature is not so simple as predicted by
a simple theory of thermochemistry.
In fact, it was not long before 
an unexpected detection of CO fundamental band at
4.6~$\mu$m in Gl~229B, classified as T6,
 was reported by \citet{Noll97} and \citet{Oppenheimer98}. 
These pioneering observations highlighted roles of non-equilibrium
chemical processes. The idea of vertical mixing was suggested by 
\citet{Griffith99} and \cite{Saumon00}.
Because of the extreme stability of CO, the chemical time scale is much 
longer than the mixing timescale and CO is dredged up by vertical mixing
from the inner part of the atmosphere where CO is still abundant to
the surface layers where CH$_4$ dominates instead of CO.
Recently two more late-T dwarfs were found to show the CO band
by \citet{Geballe09}, confirming that the presence of CO is
a general characteristic of late-T dwarfs.

It is to be remembered that the observations so far are
limited to late-T dwarfs. Observations of the CO fundamental band in early-T
and L-type dwarfs are needed for a unified understanding of
the physical and chemical processes related to CO in the 
atmospheres of brown dwarfs. Spectroscopic observations
of the CH$_4$ and of the CO transitions in $L$ and $M$-band are always
difficult from the ground. Severe atmospheric absorption and 
limited wavelength coverage make precise analysis difficult.
Thus, there is a strong motivation 
to carry out near-infrared spectroscopy of brown dwarfs from 
space, especially in the region of 2.5--5.0~$\mu$m which remains 
the least explored so far. 

The infrared astronomical satellite \textit{AKARI} \citep{Murakami07} was
launched in 2006 February. Scientific observations under
cryogenic condition were carried out from 2006 May to
2007 August, and ended with the liquid Helium boil-off.
The satellite is equipped with a 68.5 cm cooled telescope 
and two scientific instruments covering the wavelength range of 1.8--180~$\mu$m.
One of them, the Infrared Camera \citep[IRC;][]{Onaka07} carried out
imaging and spectroscopic observations in the near- to mid-infrared
wavelength regions.
The IRC provided a unique opportunity to take spectra of
brown dwarfs in the range between 2.5 and 5.0~$\mu$m
without interference of telluric absorption from the atmosphere.
In the framework of the \textit{AKARI} Mission Programme (MP; coordinated
observations by the project team members), we have carried out 
a series of near-infrared spectroscopic observations of ultracool dwarfs.
This programme, \textit{NIRLT} (PI: I.Yamamura), aims to obtain
a set of legacy data for studies of the evolution of physical
and chemical structure of L- and T-dwarfs.
In this paper we present the initial results from this programme.

As a basis of interpretation of the spectra observed with \textit{AKARI}, 
we determine basic physical parameters of six well observed objects.
For this purpose, we apply model photospheres as in 
the analyses of ordinary stars. However, the model photospheres of 
ultracool dwarfs including brown dwarfs are by no means well established yet.
It has been discussed that dust grains in the atmosphere
play an important role in the formation of spectra of ultracool dwarfs:
especially spectra of L-type dwarfs
($T_\mathrm{eff}=$ 2200--1400 K) are strongly influenced by the continuous
opacity of dust, while spectra of even cooler T-type dwarfs
are dominated by deep molecular absorption bands \citep{Tsuji96}.
In particular, we have no definite guideline for treating
dust formation in the photosphere. Some attempts to model the photospheres of brown
dwarfs with dust clouds have been carried out by several groups and
these are discussed in detail by \citet{Helling08}.
It was shown that all models agree on the global structure of
the dusty photospheres, but predicted features such as the emergent
spectra differ considerably, depending on the 
opacity data and reflecting the extreme complexity of such objects. 

We apply the Unified Cloudy Model \citep[UCM:][]{Tsuji02,Tsuji05} 
as an example of such dusty model photospheres
to the analysis of our objects. We assume Solar metallicity 
throughout, and estimate $T_\mathrm{eff}$, $T_\mathrm{cr}$ (critical 
temperature that defines the upper boundary of the dust cloud; see 
Section~\ref{sec:predspec} for detail), and $\log g$ from the relative SEDs
(i.e. shapes of the spectra). 
 We then fit the observed and predicted spectra on an absolute scale
and estimate the radii of individual objects with the use of the
known parallaxes. Based on the final fits of the model spectra to the
observed ones, we try to interpret the new features 
 in the spectral region between 2.5 and 5.0~$\mu$m observed with 
\textit{AKARI} for the first time.

\section{Observations and Data Reduction}

The \textit{AKARI}/IRC covered its wavelength range
with three independent cameras
operated simultaneously, namely the NIR (near-infrared),
MIR-S (mid-infrared short), and MIR-L (mid-infrared long) channels.
Our observations were all carried out in
the AOT (Astronomical Observation Template) IRC04 with parameters
of ``b;Np'' \citep{Lorente08}. In this mode a grism was used to achieve
the highest available spectral resolution; the dispersion was
0.0097~$\mu$m/pixel or $R = \lambda / \Delta\lambda = 120$ at
3.6~$\mu$m \citep{Ohyama07}.
Targets were placed in the $1' \times 1'$ point source aperture 
mask to minimize the
contamination from nearby stars and background sky.
The aperture mask was only available for the NIR channel, and
the spectra obtained in this programme were limited to the near-infrared
wavelength range of 2.5--5.0~$\mu$m.

Our target list consisted of 30 brown dwarfs selected by their
expected fluxes (to be bright enough for the \textit{AKARI}/IRC instrument to
provide high-quality spectra within reasonable number of pointings)
and their spectral types (to sample various types from L to T).
A pointed observation by \textit{AKARI} allowed about 10 minutes
of exposure.
At least two observations were requested per target to obtain 
data redundancy. Due to severe visibility 
constraint of the satellite, not all requested observations were carried out.
There were 18 pointing opportunities for 11 objects in the cryogenic mission
phase. A summary of the observed targets is given in Table~\ref{tbl:target},
and the observation record is presented in Table~\ref{tbl:obslog}.
The tables include observations which were not successful, for completeness.

\begin{deluxetable}{lccllllll}
   \tabletypesize{\footnotesize}
   \rotate
   \tablewidth{0.82\vsize}
   \tablecaption{Summary of the targets in the current programme\label{tbl:target}}
   \tablehead{
      \colhead{Object Name} &  \colhead{R.A.(J2000)} & \colhead{Dec.(J2000)} &
      \colhead{Sp.Type} & \colhead{[$J$]} & \colhead{[$H$]} & \colhead{[$K$]} & 
      \colhead{[$L'$]}  & \colhead{ref.}
   }
\startdata
\object[2MASS J04151954-0935066]{2MASS J04151954$-$0935066}    &  04:15:19.54 & $-$09:35:06.6 & T8    & 15.32 & 15.70 & 15.83 & 13.28    & 1,3\\
\object[SDSS J053951.99-005902.0]{SDSS J053951.99$-$005902.0}  &  05:39:52.00 & $-$00:59:01.9 & L5    & 13.85 & 13.04 & 12.40 & 11.32    & 1,3\\
\object[2MASS J05591914-1404488]{2MASS J05591914$-$1404488}    &  05:59:19.14 & $-$14:04:48.8 & T4.5  & 13.57 & 13.64 & 13.73 & 12.14    & 1,3\\
\object[SDSS J083008.12+482847.4]{SDSS J083008.12$+$482847.4}  &  08:30:08.25 & $+$48:28:48.2 & L9    & 15.22 & 14.40 & 13.68 & 11.98    & 1,3\\
\object[2MASS J12171110-0311131]{2MASS J12171110$-$0311131}    &  12:17:11.10 & $-$03:11:13.1 & T7.5  & 15.56 & 15.98 & 15.92 & 13.96    & 1,3\\
\object[GD 165B]{GD 165B}                                      &  14:24:39.09 & $+$09:17:10.4 & L3    & 15.64 & 14.75 & 14.09 & 12.93    & 1,3\\
\object[SDSS J144600.60+002452.0]{SDSS J144600.60$+$002452.0}  &  14:46:00.61 & $+$00:24:51.9 & L5    & 15.56 & 14.59 & 13.80 & 12.54\tablenotemark{a}    & 1,3\\
\object[2MASS J15232263+3014562]{2MASS J15232263$+$3014562}    &  15:23:22.63 & $+$30:14:56.2 & L8    & 15.95 & 15.05 & 14.35 & 12.86    & 1,5\\
\object[2MASS J17114573+2232044]{2MASS J17114573$+$2232044}    &  17:11:45.73 & $+$22:32:04.4 & L6.5  & 17.09 & 15.80 & 14.73 & N/A      & 2\\
\object[SDSS J175032.96+175903.9]{SDSS J175032.96$+$175903.9}  &  17:50:32.93 & $+$17:59:04.2 & T3.5  & 16.14 & 15.94 & 16.02 & 14.95\tablenotemark{b}& 1,4\\
\object[eps Ind B]{$\epsilon$ Ind Ba$+$Bb}                     &  22:04:10.52 & $-$56:46:57.7 & T1+T6 & 11.91 & 11.31 & 11.21 & ~~9.97\tablenotemark{b}& 2,4\\

\enddata
\break
   \tablerefs{
      [$J,H,K$] (1) \citet{Knapp04}, 
         (2) 2MASS Point Source Catalog, 
         (3) \citet{Golimowski04} \break
      [$L'$] (4) \citet{Patten06}, (5) \citet{Leggett02}
      }
   \tablenotetext{a}{Estimated magnitude.}
   \tablenotetext{b}{\textit{Spitzer}/IRAC 3.5~$\mu$m band magnitude.}
   \end{deluxetable}

The standard software toolkit {\it IRCSPEC\_RED} \citep{Ohyama07}
was used for the data reduction\footnote{The software is available
from the \textit{AKARI} observers' site;
\url{http://www.ir.isas.jaxa.jp/AKARI/Observation/}}. The processing
was performed in January 2008 with the pre-released version
of the toolkit, which should be equivalent with
the public version ``20080528''. We followed the standard reduction
recipe. Fine tuning of the on-source / off-source mask on
the spectral images was adopted to obtain the maximum signal
and minimum contamination.
Corrections of instrumental effects and
wavelength / flux calibrations were all done automatically
in the toolkit, and we did not observe any calibration sources
by ourselves. 
Since our observations were in principle slitless spectroscopy,
the major source of the wavelength error was the determination
of the reference point. The typical error was 0.5 pixel of the
detector or $\sim 0.005$~$\mu$m \citep{Ohyama07}, 
but could be larger on some occasions. 
We applied a small correction ($-0.008$~$\mu$m) to 
2MASS J152322$+$3014 by comparing 
the position of CH$_4$ $Q$-branch feature
with other objects and with the synthesized spectrum.
The overall flux calibration error is 10 per cent in the middle
of the wavelength range and 20 per cent at the short / long edges.
See \citet{Ohyama07} for more details about data reduction and
calibration.

\begin{table*}
  \caption{Observation log}\label{tbl:obslog}
  \begin{center}
    \begin{tabular}{lccl}
      \hline
      \hline
Object Name         &  Date       &  OBSID       &  Remarks \\
      \hline
2MASS J041519$-$0935  &  2007-02-18 &  1720005-001 &  Ghosting \\
2MASS J041519$-$0935  &  2007-02-18 &  1720005-002 &  Ghosting \\
2MASS J041519$-$0935  &  2007-08-23 &  5125080-001 &  Re-observation \\
2MASS J041519$-$0935  &  2007-08-24 &  5125081-001 &  Re-observation \\
SDSS  J053951$-$0059  &  2006-09-17 &  1720009-001 &  \\
2MASS J055919$-$1404  &  2006-09-22 &  1720006-001 &  \\
2MASS J055919$-$1404  &  2006-09-22 &  1720008-001 &  \\
SDSS  J083008$+$4828  &  2006-10-20 &  1720007-001 &  \\
SDSS  J083008$+$4828  &  2006-10-21 &  1720007-002 &  \\
2MASS J121711$-$0311  &  2007-06-26 &  1720068-001 &  Too faint \\
GD165B                &  2007-07-24 &  1720074-001 &  Data lost \\
SDSS J144600$+$0024   &  2007-08-02 &  1720072-001 &  \\
eps Ind Ba$+$Bb       &  2006-11-02 &  1720003-001 &  \\
eps Ind Ba$+$Bb       &  2006-11-02 &  1720004-001 &  \\
2MASS J152322$+$3014  &  2007-01-26 &  1720002-001 &  \\
2MASS J171145$+$2232  &  2007-03-05 &  1720001-001 &  Too faint \\
SDSS  J175032$+$1759  &  2007-03-17 &  1720050-001 &  Confusion \\
SDSS  J175032$+$1759  &  2007-03-17 &  1720050-002 &  Confusion \\
      \hline
    \end{tabular}
  \end{center}
\end{table*}

\section{Near-Infrared Spectra of Brown Dwarfs}\label{sec:nirspec}

\begin{figure*}[!ht]
  \begin{center}
   \resizebox{0.65\hsize}{!}{
       \includegraphics{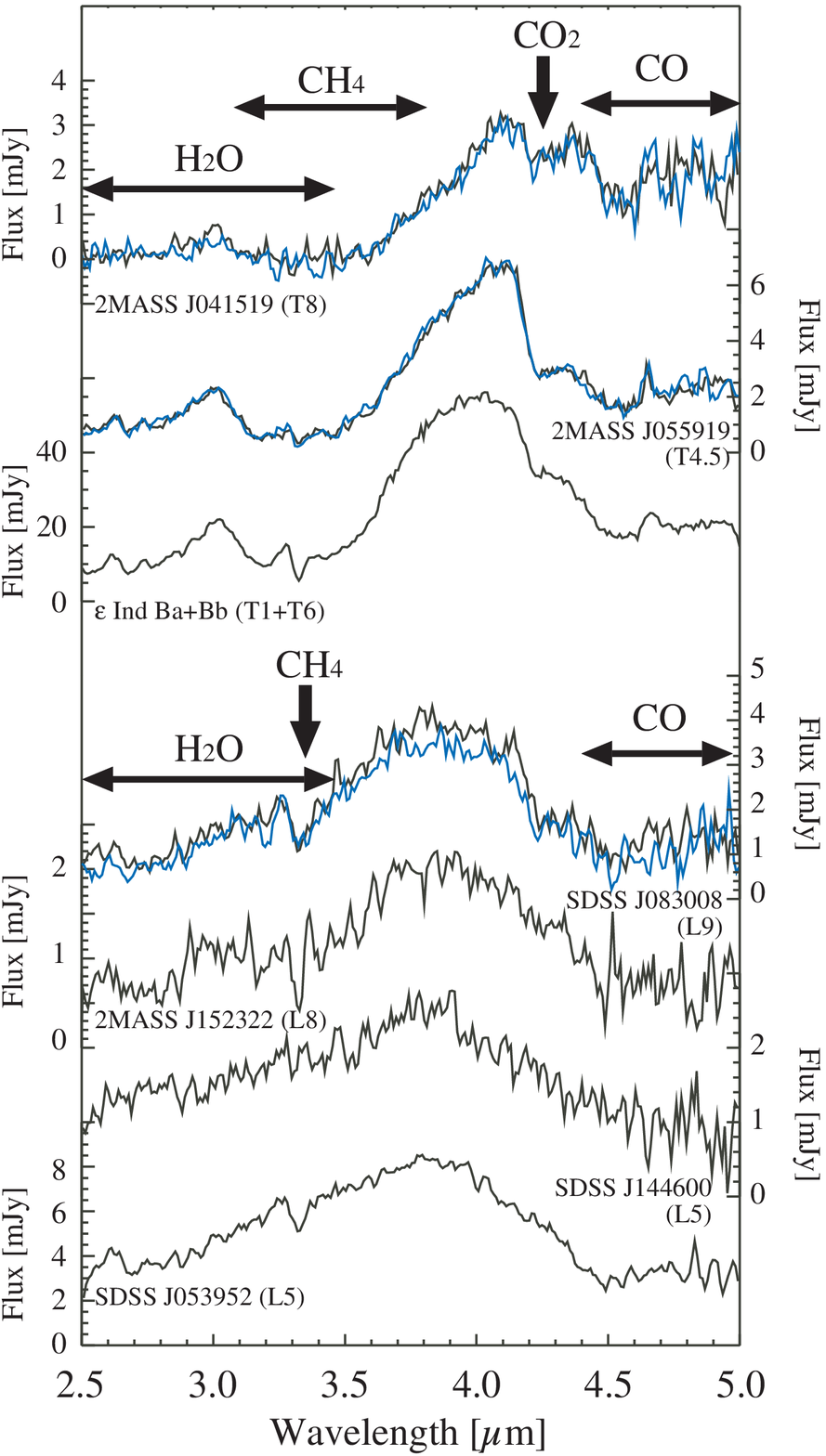}
   }
  \end{center}
  \caption{The NIR spectra of brown dwarfs obtained by the \textit{AKARI}/IRC.
   The spectra are ordered in the sequence of their spectral types
   from bottom (L5) to top (T8). When two observations were made for
   an object, the data were processed independently and the
   second spectrum is indicated in blue. The difference between the two
   observations represents the practical errors.
   Positions of major molecular bands are indicated.
}\label{fig:ircspec}
\end{figure*}

Figure~\ref{fig:ircspec} shows the observed \textit{AKARI} spectra of brown dwarfs
in the sequence of their spectral types from L (bottom) to T (top). 
When an object was observed twice, the two data sets were processed 
independently and are plotted
in different colors. In general, the two spectra are consistent
with each other, demonstrating the high quality and reliability
of the data.

The overall SEDs of the observed spectra agree well with 
the predictions from the previous photometry and theoretical works.
The spectra of L-type dwarfs are rather smooth and featureless,
mildly peaked at 3.8~$\mu$m as a result of
broad absorption bands of H$_2$O at shorter wavelengths and CO at
longer wavelengths.
On the other hand, spectra of T-type dwarfs are dominated
by deep molecular absorption features. As expected, 
we detect CH$_4$ and H$_2$O molecular bands.
The bands deepen as the spectral type evolves.
The band shapes are heavily distorted due to 
saturation in the latest object, 2MASS J041519$-$0935 (T8).

\begin{figure}[!ht]
  \begin{center}
   \resizebox{0.9\hsize}{!}{
       \includegraphics[60,150][450,565]{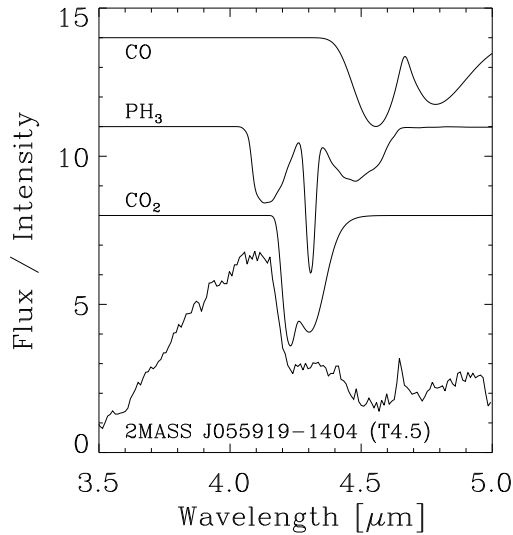}
   }
  \end{center}
\caption{The spectrum of 2MASS J055919$-$1404 is compared with
model spectra of CO$_2$, PH$_3$, and CO. The model spectra
are calculated by a simple plane-parallel configuration with
column densities and excitation temperatures of; 
$10^{18}$ cm$^{-2}$ and 700 K for CO$_2$, 
$10^{19}$ cm$^{-2}$ and 700 K for PH$_3$, and
$10^{18}$ cm$^{-2}$ and 1000 K for CO, respectively.
The CO$_2$ and CO data are taken from HITEMP database
\citep{Rothman97}, and PH$_3$ data from HITRAN2008
\citep{Rothman09}. The model spectra are normalized, and
scaled and shifted for comparison.
}\label{fig:co2}
\end{figure}

The most remarkable finding in the current 
\textit{AKARI} spectra is a detection
of the CO$_2$ fundamental stretching-mode band at 4.2~$\mu$m
in the late-L and T-type dwarfs. 
The identification of this band is demonstrated in Figure~\ref{fig:co2},
in which the spectrum of 2MASS J055919$-$1404 is compared with 
synthesized spectra of CO$_2$. Spectra of CO and PH$_3$ are
also presented for reference.
The shape of the band profile, especially a sharp edge around
4.2~$\mu$m confirms the identification of the feature as CO$_2$.

This is the first detection of CO$_2$
in brown dwarfs, since the wavelength is inaccessible
from the ground or even from airborne observations due to
huge opacity of the molecule itself in the telluric atmosphere. 
The band is most apparent in the T4.5 dwarf 2MASS J055919$-$1404,
and is clearly recognized in 2MASS J041519$-$0935 (T8) and
$\epsilon$ Ind Ba+Bb (T1+T6).
It is seen in the L9 dwarf SDSS J083008$+$4828 and very
marginally in the L8 dwarf 2MASS J152322$+$3014. We suspect that
the formation of the molecule takes place in a relatively
low temperature environment.

Another important result from our observations is that
we see the CO fundamental band in the 4.4--5.0~$\mu$m region
in all of our samples including the coolest T8 dwarf, 2MASS J041519$-$0935.
The detection of the CO molecule in a late-T dwarf was first
reported by \citet{Noll97} and \citet{Oppenheimer98} in the T6 dwarf
Gl~229B. Two more objects have been added recently by \citet{Geballe09};
2MASS J09373487$+$2931409 (T6) and Gl~570D (T7.5).
The presence of the CO band is also suggested from the \textit{Spitzer}
3.5--7.9~$\mu$m photometry \citep{Patten06, Leggett07}.
Our \textit{AKARI} spectra provide direct and indubitable evidence
of the molecule and strongly support the idea that the presence
of the CO band is a common feature in the late-T dwarfs.

We also detect the CH$_4$ molecule in SDSS J053951$-$0059 (L5),
confirming the previous report by \citet{Noll00} that 
the molecule is already present in the atmosphere of L5 dwarfs.
The feature is not clear in another L5 source
SDSS J144600$+$0024, partly because of its relatively low S/N.
The band develops in the L8 dwarf, 2MASS J152322$+$3014,
and starts showing the $P$- and $R$-branches.

Because of moderate spectral resolution it is difficult to identify
any more molecules in the spectra. 

\subsection{Comments on individual objects}

\subsubsection{\it SDSS J053951$-$0059}
\citet{Fan00} selected this object as a high-$z$ quasar candidate
from the SDSS commissioning image and found
it to be an L5 dwarf through follow-up spectroscopy.
Due to its relatively high flux level ($\sim 8$ mJy at 3.8~$\mu$m)
the quality of the obtained spectrum is good. We can clearly
recognize the presence of CH$_4$ $Q$-branch absorption at 3.3~$\mu$m
on the otherwise rather smooth spectrum.

\subsubsection{\it SDSS J144600$+$0024}
This L5 object was nominated as a brown dwarf by \citet{Geballe02}.
Because of relatively low S/N compared to another L5
object SDSS J053951$-$0059 above, we do not recognize the CH$_4$
$Q$-branch dip in the spectrum of this source, while
the overall SEDs are consistent with each other.

\subsubsection{\it 2MASS J152322$+$3014}
\citet{McLean00} found that this source is of the latest L-type
and assigned it as L8/L9. It was confirmed as L8 by \citet{Geballe02}.
This source, together with SDSS J144600$+$0024, is one of the faintest
objects among our current dataset, and the quality of the spectrum 
is not excellent.
However, we clearly see a more developed CH$_4$ band than in
the L5 sources. The CO$_2$ band is marginally detected.

\subsubsection{\it SDSS J083008$+$4828}
\citet{Geballe02} classified this object as L8--T0, or L9 as
an average. It is on the border of L/T spectral types and is
a good example to investigate the transition from L to T.
There are rather remarkable changes in the spectral features
between this source from the previous L8 source 2MASS J152322$+$3014.
We see that H$_2$O and CO absorption bands
become more prominent than earlier L-type sources.
The CH$_4$ $Q$-branch is very clear, but $P$- and $R$-branches are
less obvious. The CO$_2$ band is clearly seen.

\subsubsection{\it $\epsilon$ Ind Ba$+$Bb}
This object was identified by \citet{Scholz03} as a companion
to the K5V star, $\epsilon$~Ind~A. Because of its proximity to the Sun
(3.6 pc), it is the brightest among our observed targets.
We observed the object twice within a day
but found that one observation (OBSID=1720003-1) 
was significantly degraded by charged particle hits 
before and during the exposure.
The other spectrum (OBSID=1720004-1) shown in Figure~\ref{fig:ircspec}
clearly exhibits the four major molecular bands;
CH$_4$, H$_2$O, CO and CO$_2$. This object is
in fact a binary system of T1 and T6 dwarfs, with $K$-band
magnitudes of 14 and 16, respectively \citep{MacCaughrean04}.
\textit{AKARI}'s spectrum is a mixture of the two spectral types and
we will not use it for quantitative analysis.

\subsubsection{\it 2MASS J055919$-$1404}
The object was identified as a `warm' T-dwarf by \citet{Burgasser00}.
From a near-infrared spectrum in 0.9--2.3~$\mu$m they concluded that
the source was near the L/T transition border. \citet{Burgasser06b}
classified the object as T4.5.
The \textit{AKARI} spectra of this dwarf show very deep CO and CO$_2$
absorptions beyond 4.0~$\mu$m, together with CH$_4$ and H$_2$O
bands below 3.8~$\mu$m. As a result the spectra show a peak at
4~$\mu$m. A sharp peak, probably the CO band center appears 
very clearly between the 
$R$- and $P$-branches. Our observations are consistent with
the photometry by \citet{Leggett02} that the $M'$ band flux of this
source is fainter by a factor of three than that expected
if CO/CH$_4$ abundance follows thermal equilibrium.
Our spectra confirm that CO is the major source of opacity
in the wavelength region.

\subsubsection{\it 2MASS J041519$-$0935}
This object was first identified as a T-type brown dwarf
by \citet{Burgasser02} based on near infrared spectroscopy.
The object has been designated
a ``standard'' T8 source by \citet{Burgasser06b}.
It is one of the latest spectral-type brown dwarf,
and is the coolest object in our sample.
The first observations of this target with \textit{AKARI} were 
carried out in 2007 February. However, the data were 
considerably contaminated by ghosting from a nearby 
bright star.
We reobserved the object a half year later in 2007 August
and obtained a clean spectrum.

The spectra exhibit deep H$_2$O and CH$_4$ absorption bands,
which are almost completely saturated at the wavelengths
below 3.5~$\mu$m. On the other hand, the object has sufficient flux in
the longer wavelength range, where we clearly identify the CO and CO$_2$ bands.
Two independent observations show almost identical
shape in this wavelength range, strengthening the reality of the detections.

\section{Comparisons of the Observed and Model Spectra}\label{sec:ucmfits}
  In this section, we examine to what extent the newly
observed spectra of the 2.5--5.0~$\mu$m region can be explained 
with the model spectra and what remains unexplained 
with the present models of the photospheres of cool dwarfs. 
For this purpose, we apply our Unified Cloudy Model (UCM),
a brief description of which is given in Section~\ref{sec:predspec}.
We assume local thermodynamical equilibrium (LTE) throughout
this section. Then we outline our method of analysis with some
examples in Section~\ref{sec:obsvsucm} and the results of our 
analysis in Section~\ref{sec:ucmfitresults}.
The physical parameters estimated by our analysis are summarized
in Section~\ref{sec:params}.
Finally, we examine the errors and limitation of
our analysis based on the UCM in Section~\ref{sec:errors}.

\subsection{Predicted spectra of dusty dwarfs}\label{sec:predspec}

Generally, stellar spectra can be interpreted in terms
of $T_\mathrm{eff}$, $\log g$, chemical composition, and micro-turbulent
velocity. However, a new feature in the spectra
of brown dwarfs, compared with usual stellar spectra, is
the effect of dust clouds formed in the photosphere.
If the properties of dust clouds can be defined uniquely from
the four basic parameters considered above, the spectra of
brown dwarfs could eventually be interpreted in terms of these
four parameters, namely $T_\mathrm{eff}$, $\log g$, chemical composition, and 
micro-turbulent velocity. At present, the dependence of the dust cloud
properties on the four basic parameters is unknown, because the formation
and disappearance of dust clouds in the photospheres of brown dwarfs is not yet well understood.

\begin{figure}[!ht]
  \begin{center}
   \resizebox{1.0\hsize}{!}{
       \includegraphics{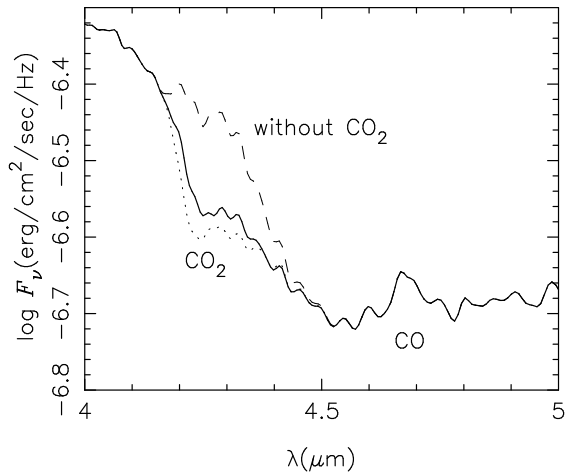}
   }
  \end{center}

\caption{ The predicted spectrum of a UCM with $T_\mathrm{eff} =$ 1300 K,
$T_\mathrm{cr} = $1800 K, $\log g = 4.5$, $\xi_\mathrm{micro} = 
$1 km\,s$^{-1}$, and Solar metallicity, showing the effect of the 
CO$_{2}$ 4.2~$\mu$m band based on the line
list (solid line) and on the band model opacity (dotted line)
compared with the case without CO$_{2}$ lines (dashed line).
}\label{fig:ucmco2}
\end{figure}

On the other hand, some observations indicate that the properties of
the dust clouds do not necessarily depend uniquely on such
basic parameter as $T_\mathrm{eff}$. For example, infrared colors such as
$J-K$ plotted against $T_\mathrm{eff}$ show a large variation at a fixed
$T_\mathrm{eff}$ \citep[e.g.][]{Marley05, Tsuji05}. Since these infrared 
colors depend directly on the properties of the dust clouds, this fact
implies that the properties of the dust clouds should be different 
even for the same $T_\mathrm{eff}$. Also, one important observational result 
is that $T_\mathrm{eff}$ shows little change over the spectral types 
between about L5 and T5 \citep[e.g.][]{Golimowski04, Nakajima04}. 
In other words, the characteristics of the spectra used as
signatures of the spectral types are quite different even for the same 
$T_\mathrm{eff}$ and hence they are not necessarily be 
determined by $T_\mathrm{eff}$. This result implies that the basic 
features of the spectra of brown dwarfs 
should be determined by some additional parameter(s) other than
$T_\mathrm{eff}$. Again dust should play a major role in this respect as in 
the case of the infrared colors noted above. 

Thus, a major problem is how to consider the effect of dust in
the interpretation and analysis of the spectra of brown dwarfs.
For this purpose, we apply our Unified Cloudy Model (UCM) in
which the dust forms at a layer where temperature is equal to the
condensation temperature, $T_\mathrm{cond}$, but disappears at somewhat lower 
temperature which we refer to as the critical temperature, 
$T_\mathrm{cr}$. 
At this temperature, the radius of the dust grain $r_\mathrm{gr}$ reaches 
the critical radius $r_\mathrm{cr}$, where the Gibbs energy of condensation
is maximum. Then, at this point, dust starts to grow larger 
and larger, and will precipitate from the gaseous mixture. 
For this reason, dust disappears in the layers with $T < T_\mathrm{cr}$, 
and will exist only in the layers of ${T_\mathrm{cr} < T < T_\mathrm{cond}}$. 
Thus dust is present in the layers whose photospheric temperatures 
are within the above range, and this means that a thin dust cloud forms in the 
photosphere of brown dwarfs,  as  discussed in more detail elsewhere
\citep{Tsuji01,Tsuji02}. It is to be noted that $T_\mathrm{cond}$
is defined by the thermodynamical data and is hence fully controlled
by $T_\mathrm{eff}$ and $\log g$ (see e.g. Figure~3 in \citet{Tsuji02}). 
On the other hand, $T_\mathrm{cr}$ is not predictable by any physical
theory at present. Instead, $T_\mathrm{cr}$ is introduced as a free parameter
in our UCM, and it serves as a measure of the thickness of the dust 
cloud (the dust cloud is thicker if the deviation of
$T_\mathrm{cr}$ from  $T_\mathrm{cond}$ is larger). 

With the introduction
of $T_\mathrm{cr}$, the UCM is characterized by the five parameters including
$T_\mathrm{cr}$ in addition to the usual four parameters, 
$T_\mathrm{eff}$, $\log g$, chemical composition, and
micro-turbulent velocity.
From observations, we showed that at least one additional
parameter is needed to describe the observed characteristics of brown dwarfs, 
and we now propose to identify $T_\mathrm{cr}$ as the  additional parameter
required from observations.

   A new feature in the \textit{AKARI} observations discussed in
Section~\ref{sec:nirspec} is the
detection of the CO$_{2}$ band at 4.2~$\mu$m. Although CO$_{2}$ is included 
in the construction of the UCM as well as in our evaluation of the spectra
with the use of the band model opacity \citep[e.g. Appendix of][]{Tsuji02}, 
we recompute the spectra with the CO$_{2}$ line list from the HITEMP 
database \citep{Rothman97} in view of the detection of the CO$_{2}$ band 
in several objects. As an example,
a predicted spectrum including CO$_{2}$ based on the line list (solid line) 
is compared with that without CO$_{2}$ (dashed line)
in Figure~\ref{fig:ucmco2} for a UCM with $T_\mathrm{eff} =$ 1300 K, 
$T_\mathrm{cr} =$ 1800 K, $\log g = 4.5$, 
$\xi_\mathrm{micro} =$ 1 km\,s$^{-1}$, and Solar abundance.
For comparison, the result based on the band model opacity 
is shown by the dotted line. We use the band model opacity during the
iterations of model construction but apply the line list in all the
computation of the spectra. 
So far, CO$_{2}$ was largely neglected in spectroscopy and model
photospheres of cool dwarfs, but the result shown in
Figure~\ref{fig:ucmco2} reveals that the effect of CO$_{2}$ absorption
is significant
even within the framework of the LTE analysis. Since carbon atoms
are mostly consumed in CO and/or CH$_{4}$ in cool dwarfs, CO$_{2}$ cannot
be very abundant, but rather large $f$-values of CO$_{2}$ transitions make the
molecule an important absorber in the spectra of cool dwarfs. 

The major spectral feature in the 2.5--5.0~$\mu$m region is the CH$_{4}$
fundamental band at 3.3~$\mu$m, for which we apply the line list by
Freedman (2005, private communication) discussed in detail by 
\citet{Freedman08}. 
Certainly the CH$_{4}$ line lists including the high excitation lines 
are not yet satisfactory as discussed before \citep{Tsuji05}, 
but we believe that Freedman's list is the best one currently available
for the fundamental band. Other line lists included are: 
H$_2$O \citep{Partridge97}, CO \citep{Guelachivili83, Chackerian83}, 
OH \citep{Jacquinet-Husson99}, CN \citep{Cerny78, Bauschlicher88}, and 
SiO \citep{Lavas81, Tipping81}. Also, NH$_{3}$, PH$_{3}$, and H$_{2}$S 
are considered on the basis of the band models.
With these spectroscopic data, the spectra between 2.5 and 5.0~$\mu$m
for our UCMs are evaluated at a resolution of 0.01 
cm$^{-1}$ and convolved with the slit function of FWHM = 3000\,km\,s$^{-1}$, 
which is about the resolving power of the \textit{AKARI} 
spectrometer. Note that the observed spectra shown in 
Figures~\ref{fig:fit0830},~\ref{fig:fit0559},~\&~\ref{fig:fitall}
are smoothed with $R=100$, and that the error weighted
average is taken if two valid spectra are available.

\subsection {Method of analysis and some examples}
\label{sec:obsvsucm}

We restrict our analysis to the spectral region between 2.5 and
5.0~$\mu$m, which we have observed completely for the first time. 
Although analyses of different spectral regions do not necessarily 
provide the same answer as to what are the best parameters of
the photosphere, as shown in detail by \citet{Cushing08}, 
we restrict our analysis to the 2.5--5.0~$\mu$m region for this
very reason. In fact, for our purpose stated at 
the beginning of this Section, it is meaningless to find a solution that
provides a good fit to other  regions but  not to
the 2.5--5.0~$\mu$m region.  We return to this problem in 
Section~\ref{sec:errors}. 

We assume that the objects we have observed are all of Solar
metallicity \citep{Anders89, AllendePrieto02} and that the micro-turbulent 
velocity is near solar (1~km\,s$^{-1}$) throughout. 
Then, major parameters 
that define the characteristics  of a spectrum are $T_\mathrm{eff}$, 
$T_\mathrm{cr}$, and $\log g$. In fitting the observed spectrum
with the predicted one, we refer to the empirical data 
summarized in Table~\ref{tbl:knowndata}. We start with the parallaxes 
by \citet{Vrba04} 
which covered  our full sample. We notice that their values agree quite 
well with other results in general. We choose $T_\mathrm{eff}$ 
to be $T_\mathrm{eff}^{0}$ close to the 
so-called empirical $T_\mathrm{eff}$ suggested by \citet{Vrba04}
 (Table~\ref{tbl:knowndata}) and examine the cases of
$T_\mathrm{eff}$ = $T_\mathrm{eff}^{0}$ and $T_\mathrm{eff}^{0} \pm 100$\,K. 
As a result, we examine 
three values of $T_\mathrm{eff}$ near the empirical $T_\mathrm{eff}$. But
this choice of trial $T_\mathrm{eff}$'s is not necessarily successful
as described below (e.g. a case of 2MASS J055919$-$1404 to be discussed 
in Section~\ref{sec:obsvsucm}). This fact suggests that the so-called 
empirical $T_\mathrm{eff}$ cannot be fully
realistic as will be discussed in Section~\ref{sec:teffr}. 
We also consider three cases of $T_\mathrm{cr} =$ 1700, 1800, \& 1900\,K
and three values of $\log g =$ 4.5, 5.0, 5.5.
Thus, we consider $3 \times 3 \times 3 = 27$ combinations of these parameters 
($T_\mathrm{eff}$, $T_\mathrm{cr}$, and $\log g$), 
and select the case that shows a best fit to the overall shape of
the 2.5--5.0~$\mu$m SED.

\begin{figure*}
  \begin{center}
   \resizebox{0.7\hsize}{!}{
       \includegraphics{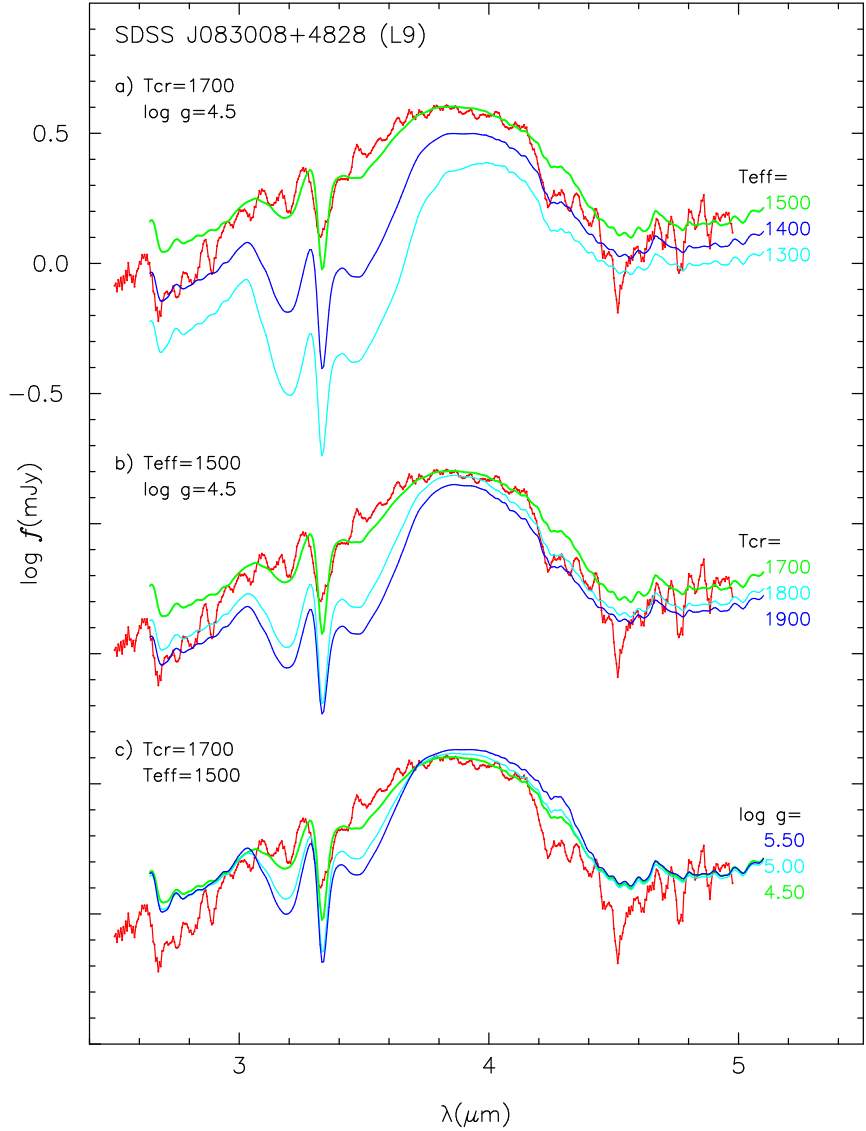}
   }
  \end{center}
\caption{ The observed spectrum of SDSS J083008$+$4828 (L9) is compared
with the predicted spectra based on the UCM. 
The best fit obtained for ($T_\mathrm{eff}$, $T_\mathrm{cr}$, $\log g$) = 
(1500, 1700, 4.5) is shown in each plot. Flux is adjusted by
shifting the model spectra vertically. The same amount of shift is 
applied to the other predicted spectra. 
Hence no attempt is made to fit them to the observed
spectrum, but are shown here for comparison purposes. 
a) Effect of $T_\mathrm{eff}$ under the fixed values of $T_\mathrm{cr}$=1700\,K and $\log g = 4.5$.
b) Effect of $T_\mathrm{cr}$ under the fixed values of $T_\mathrm{eff}$=1500\,K and $\log g = 4.5$.
c) Effect of $\log g$ under the fixed value of $T_\mathrm{cr}$=1700\,K and $T_\mathrm{eff} =$ 1500\,K.
}\label{fig:fit0830}
\end{figure*}

 We discuss   the procedure outlined above for the case of
SDSS J083008$+$4828 as an example.
The empirical $T_\mathrm{eff}$ of this L9 dwarf is 1327\,K 
(Table~\ref{tbl:knowndata}), but the CH$_{4}$ 3.3~$\mu$m band 
is too weak for such a low value of $T_\mathrm{eff}$. Therefore we start 
our trials with $T_\mathrm{eff} =$ 1300\,K and extend then to higher 
$T_\mathrm{eff}$, rather than to examine the cases of $T_\mathrm{eff} = 1300
\pm 100$\,K. It is rather difficult to find a good fit in this object, 
but one possible solution is obtained for ($T_\mathrm{eff}$, 
$T_\mathrm{cr}$, $\log g$) = (1500, 1700, 4.5) from our 27 trial models. 
The effect of $T_\mathrm{eff}$ along 
with the possible best case is shown in Figure~\ref{fig:fit0830}a: 
only the case of $T_\mathrm{eff}$=1500K provides a reasonable fit to the $P$- and
$R$-branches of CH$_{4}$ fundamentals, although the predicted $Q$-branch is a
bit stronger than that in the observed spectrum. 

The effect of $T_\mathrm{cr}$ at fixed values of $T_\mathrm{eff} =$ 1500 K and 
$\log g = 4.5$ is illustrated in Figure~\ref{fig:fit0830}b. Although the fit 
to the observed CO$_{2}$ 4.2~$\mu$m feature can be improved by higher 
values of $T_\mathrm{cr}$ (i.e. with thinner dust cloud), the fit to the 
CH$_{4}$ 3.3~$\mu$m feature appears to be worse. The effect of $\log g$ is 
rather modest as shown in Figure~\ref{fig:fit0830}c, but the lowest gravity 
of $\log g = 4.5$ provides better fits both to the CH$_{4}$ 3.3~$\mu$m 
and CO$_{2}$ 4.2~$\mu$m bands than the higher values. 
The final fit we find, however, is not so good around 2.7, 4.2, and
4.5~$\mu$m and we will discuss these wavelength regions in 
Section~\ref{sec:ucmfitresults}.

\begin{figure*}
  \begin{center}
   \resizebox{0.7\hsize}{!}{
       \includegraphics{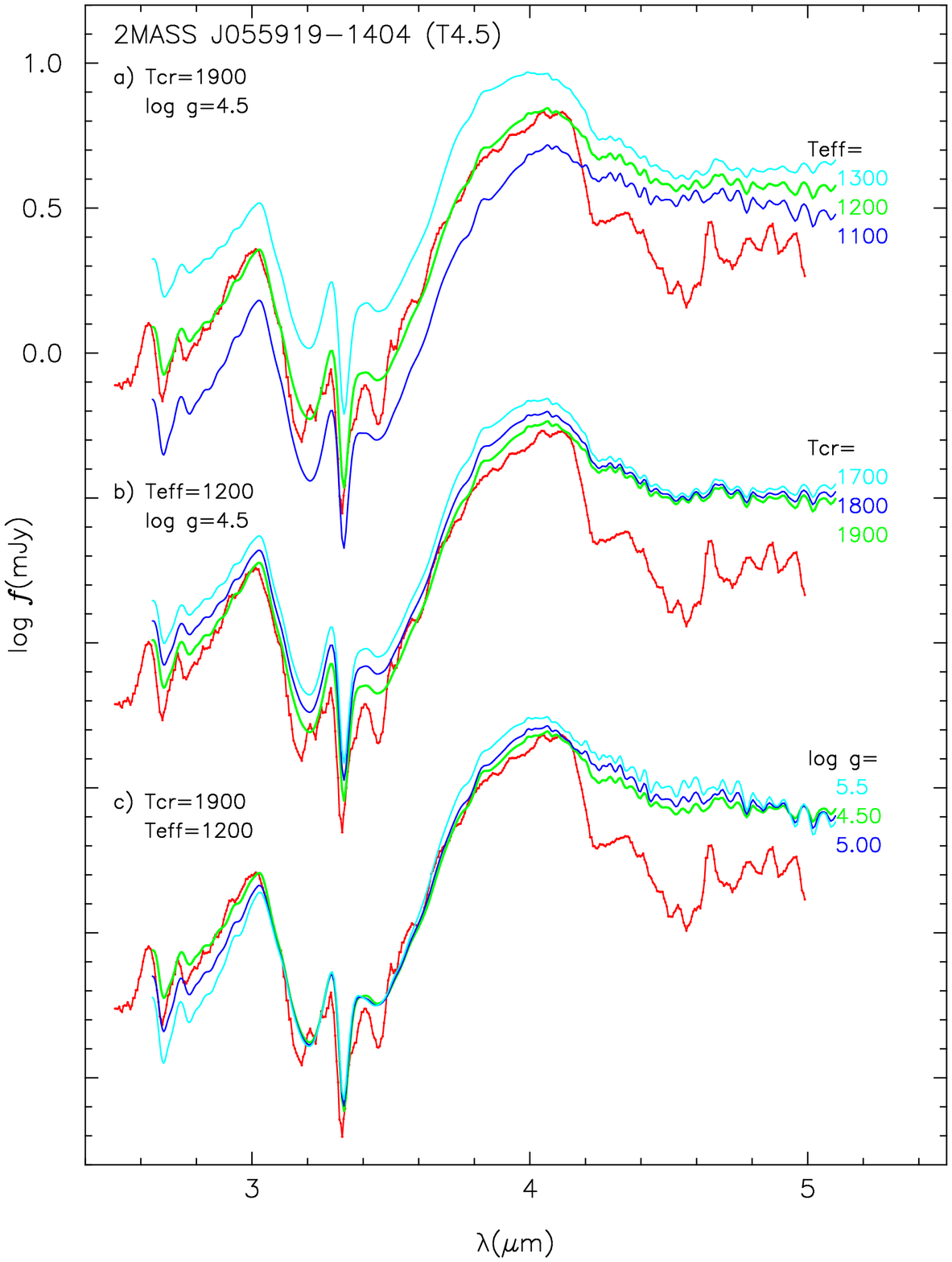}
   }
  \end{center}
\caption{ The observed spectrum of 2MASS J055919$-$1404 (T4.5) is compared
with predicted spectra based on the UCM. The best fit is obtained
for ($T_\mathrm{eff}$, $T_\mathrm{cr}$, $\log g$) = (1200, 1900, 4.5). 
See the caption of Figure~\ref{fig:fit0830} as for other details. 
a) Effect of $T_\mathrm{eff}$ under the fixed values of $T_\mathrm{cr}$=1900\,K and $\log g = 4.5$.
b) Effect of $T_\mathrm{cr}$ under the fixed values of $T_\mathrm{eff}$=1200 K and $\log g = 4.5$.
c) Effect of $\log g$ under the fixed value of $T_\mathrm{cr}$=1900 K and 
$T_\mathrm{eff} =$ 1200 K.
}\label{fig:fit0559}
\end{figure*}

Another example is the case of 2MASS J055919$-$1404.
The empirical $T_\mathrm{eff}$ of this T4.5 dwarf is 1469 K 
(Table~\ref{tbl:knowndata}), and we, at first, examined the cases of 
$T_\mathrm{eff} =$ 1400 $\pm$ 100 K. It was difficult
to find a good fit, but we found that the case of 
$T_\mathrm{eff}$ = 1300 provides a relatively close fit to some 
features.   
However, an anonymous  referee called our attention to a recent
result by \citet{Cushing08} who obtained  $T_\mathrm{eff}$ = 1200K for
this object, and suggested to examine this case. Although this case  
differs by a large value ($\approx 300$\,K!) from the empirical value, 
we now include $T_\mathrm{eff}$ = 1200\,K in our trial and find that this
case  actually provides a better fit as shown in Figure~\ref{fig:fit0559}a.
For example, the relatively strong CH$_{4}$ fundamental band can be
better explained by $T_\mathrm{eff}$ = 1200 K than by
$T_\mathrm{eff}$ = 1300 K. However, the CO$_{2}$ 4.2~$\mu$m feature 
as well as the strong absorption at the positions of CO fundamentals
cannot be fitted at all. We will return to this problem in 
Section~\ref{sec:ucmfitresults}. 

If we assume values of $T_\mathrm{cr}$ lower than 1900 K, the fits to the 
CH$_{4}$ 3.3~$\mu$m band tend to become worse while those to the CO$_{2}$ and 
possible CO bands show little change as seen in Figure~\ref{fig:fit0559}b. 
The $J-K$ color near zero (Table~\ref{tbl:knowndata}) also suggests
that the dust cloud is rather thin, consistent with the
high value of $T_\mathrm{cr}$. The effect of $\log g$ is examined for the fixed
values of $T_\mathrm{eff} =$ 1200\,K and $T_\mathrm{cr} =$ 1900\,K.
It is found that the spectrum  shows little change for
the change of $\log g$, but the 2.8--3.0~$\mu$m feature tends 
to fit better for the lower values of $\log g$ (Figure~\ref{fig:fit0559}c). 
We conclude that the case of ($T_\mathrm{eff}$, $T_\mathrm{cr}$, $\log g$) = 
(1200, 1900, 4.5) is the best compromise for this object.

\begin{table*}
\caption{The known observed data}\label{tbl:knowndata}
\begin{center}
\begin{tabular}{llcccc}
\noalign{\bigskip}
\hline
\noalign{\bigskip}
no. & object & sp. type$^{~a}$ & $J-K^{~b}$  &  parallax$^{~c}$ & 
$T_\mathrm{eff}^{~c}$  \\
    &        & Opt./IR &     & (marcsec)   &  (K) \\
\noalign{\bigskip}
\hline 
\noalign{\bigskip}
1 & SDSS  J053951$-$0059 & L5/L5 &  $1.45$ &  $76.12 \pm 2.17$ & 1690 ($+173$, $-137$)\\
2 & SDSS  J144600$+$0024 & L6/L5 &  $1.76$ &  $45.46 \pm 3.25$ & 1592 ($+175$, $-140$) \\
3 & 2MASS J152322$+$3014 & L8/L8 &  $1.60$ &  $57.30 \pm 3.27$ & 1330 ($+142$, $-112$) \\
4 & SDSS  J083008$+$4828 & L8/L9 &  $1.54$ &  $76.42 \pm 3.43$ & 1327 ($+140$, $-111$) \\ 
5 & 2MASS J055919$-$1404 & T4.5  & $-0.16$ &  $95.53 \pm 1.44$ & 1469 ($+153$, $-122$)\\
6 & 2MASS J041519$-$0935 & T8    & $-0.51$ & $174.34 \pm 2.76$ & ~764 (~$+88$, ~$-71$)\\
\noalign{\bigskip}
\hline
\end{tabular}
\end{center}
Ref.\\
  a) \citet{Kirkpatrick05} for L and \citet{Burgasser06b} for T\\
  b) \citet{Knapp04}\\
  c) \citet{Vrba04}\\
\end{table*}

\begin{table*}
\caption{ Basic parameters estimated from the model fittings using UCMs}\label{tbl:fit}
\begin{center}
\begin{tabular}{clccccc}
\noalign{\bigskip}
\hline
\noalign{\bigskip}
 no. & object & $T_\mathrm{eff}$\,(K) & $T_\mathrm{cr}$\,(K) & $\log g$ & 
$R/R_{J}^{~a}$ & $\Delta\,T_\mathrm{eff}$\,(K)$^{~b}$\\ 
\noalign{\bigskip}
\hline
\noalign{\bigskip}
1 & SDSS J053952$-$0059  & 1800 & 1800 & 5.5 & 0.804 & $-110$ \\ 
2 & SDSS J144600$+$0024  & 1700 & 1700 & 4.5 & 0.716 & $-108$ \\ 
3 & 2MASS J152322$+$3014 & 1500 & 1700 & 4.5 & 0.684 & $-170$ \\
4 & SDSS J083008$+$4828  & 1500 & 1700 & 4.5 & 0.700 & $-173$ \\
5 & 2MASS J055919$-$1404 & 1200 & 1900 & 4.5 & 1.175 & $-269$ \\
6 & 2MASS J041519$-$0935 & ~800 & $T_\mathrm{cond}$ & 4.5 & 0.767 & $-36$ \\
\noalign{\bigskip}
\hline
\end{tabular}
\end{center}

 a) radius $R$ relative to the Jupiter's radius $R_{J}$ (see Section~\ref{sec:ucmfitresults}).

 b) $\Delta\,T_\mathrm{eff} = T_\mathrm{eff}$\,(Table~\ref{tbl:knowndata} by empirical method) 
  - $T_\mathrm{eff}$\,(Table~\ref{tbl:fit} by our model fitting). 
\end{table*}

\subsection {Results}\label{sec:ucmfitresults}
We now extend the same procedure outlined in Section~\ref{sec:obsvsucm} 
to the other objects except for $\epsilon$ Ind Ba+Bb 
because of the composite nature of its spectrum due to its binarity.
The observed and predicted spectra can be fitted rather well
in the remaining four objects (see Figure~\ref{fig:fitall}), and thus they are more or
less easier to analyze compared to  the  cases discussed in
Section~\ref{sec:obsvsucm}. For this reason, we
only give the resulting best fits for them\footnote{The numerical 
details of the predicted spectra and the UCMs
are available from \url{http://www.mtk.ioa.s.u-tokyo.ac.jp/\~ttsuji/
export/ucmLM, ucm}, respectively.}. The parameters
that give the best fits for all six objects are summarized in Table~\ref{tbl:fit}. 
After we find the possible best model for each object, we measure the 
vertical shift in logarithmic scale between the observed spectrum and the 
predicted one based on the best model we found. 
The emergent flux from the unit surface area of the object,
$F_{\nu}$ (in units of erg\,cm$^{-2}$\,s$^{-1}$\,Hz$^{-1}$), is written as
$$\log F_\nu = \log f_\nu - 2 \log (R/d) - 23.497, $$
where $f_{\nu}$ (in unit of Jy) is the observed flux, $d$ is the
distance to the object based on the measured parallax, and $R$ is the
radius of the object. The vertical shift provides $R/d$. With the known
distance, we can estimate the radius $R$ and the result is also given 
in Table~\ref{tbl:fit}. 
The resulting fits in absolute scale are given in Figure~\ref{fig:fitall}.
 
\subsubsection {SDSS J053951$-$0059}\label{sec:fit0539}

  The empirical $T_\mathrm{eff}$ of this L5 dwarf is 1690\,K 
(Table~\ref{tbl:knowndata}), 
and we find that the case of ($T_\mathrm{eff}$, $T_\mathrm{cr}$, $\log g$) =
(1800, 1800, 5.5) shows the best fit for this object. In fact, the overall 
shape of the SED as well as the strengths of the CH$_{4}$ $Q$-branch at 
3.3~$\mu$m and the CO fundamental at 4.6~$\mu$m is reasonably fit for 
this case as shown in Figure~\ref{fig:fitall}a. It is to be noted that the 
effect of $T_\mathrm{cr}$ is of comparable importance to that of 
$T_\mathrm{eff}$.
In fact, the column densities of both dust and molecules generally increase
at the lower $T_\mathrm{eff}$ (at fixed $T_\mathrm{cr}$) and also at the 
lower $T_\mathrm{cr}$ (at fixed $T_\mathrm{eff}$). For this reason, the effects 
of $T_\mathrm{eff}$ and $T_\mathrm{cr}$ on the spectra are often difficult to 
discriminate. We have considered the effects of both $T_\mathrm{eff}$ and 
$T_\mathrm{cr}$ from the beginning, and looked for the best solution. 
The effect of $\log g$ on the spectrum is rather minor compared with the 
other parameters, but the lower gravities may be reasonably excluded from 
the overall poor fits. 

The resulting fit in absolute scale (Figure~\ref{fig:fitall}a) shows that the
spectrum observed by \textit{AKARI} can reasonably be accounted for
by our model. 

\subsubsection { SDSS J144600$+$0024 }
Although the empirical $T_\mathrm{eff}$ of this L5 dwarf is near 1600\,K 
(Table~\ref{tbl:knowndata}), we find it very difficult to assume 
$T_\mathrm{eff} < 1600$\,K since the observed CH$_{4}$ feature at 3.3~$\mu$m 
is very weak if present. Again by the procedure outlined in 
Section~\ref{sec:obsvsucm}, we find that a case of ($T_\mathrm{eff}$, 
$T_\mathrm{cr}$, $\log g$) = (1700, 1700, 4.5) 
shows the best fit for this object. For example, we see that not only
the strength of the CH$_{4}$ 3.3~$\mu$m band but also  the overall shape
of SED are best fitted with $T_\mathrm{eff} =$1700K. The effect of
$T_\mathrm{cr}$ is rather large and we can easily exclude such high values of 
$T_\mathrm{cr}$ as 1800 and 1900K. Also, the very red $J-K$ color 
(Table~\ref{tbl:knowndata}) suggests that the dust cloud is pretty 
thick and $T_\mathrm{cr}$ cannot be as high as 1900K.
It is found that the overall SED is again not sensitive to $\log g$,
but the strengths of both the CH$_{4}$ $Q$-branch at 3.3~$\mu$m and CO 
fundamental band could discriminate the best $\log g$ of 4.5 from 
the higher values of $\log g = 5.0$ and 5.5.

The resulting fit in absolute scale in Figure~\ref{fig:fitall}b shows
that the observed
spectrum of the 2.5--5.0~$\mu$m region, which possesses almost no distinct
features,  can reasonably be accounted for by our model.

\subsubsection{2MASS J152322$+$3014}\label{sec:fit1523}
 The empirical $T_\mathrm{eff}$ of this L8 dwarf is 1330K 
(Table~\ref{tbl:knowndata}), but we found that the best fit is found for
($T_\mathrm{eff}$, $T_\mathrm{cr}$, $\log g$) = (1500, 1700, 4.5).
In fact, the modest strength of the CH$_{4}$ 3.3~$\mu$m band suggests that 
$T_\mathrm{eff}$ cannot be below 1300 K, and it is only the case of 
$T_\mathrm{eff} = $ 1500\,K that provides a reasonable fit not only to the 
$Q$- but also to the $P$- and $R$-branches of the CH$_{4}$ fundamental band. 
We confirm that such a fit as shown for $T_\mathrm{cr} =$ 1700\,K can
no longer be found for $T_\mathrm{cr}$ as high as 1800--1900 K.
This result is consistent with the red $J-K$ color of 1.60 
(Table~\ref{tbl:knowndata}),
which implies that the dust cloud of this L8 dwarf is rather thick.
The strength of the CO fundamental band is almost independent of
$\log g$, but the $P$- and $R$-branches of the CH$_{4}$ fundamentals 
depend on $\log g$ and can only be explained with the low gravity of 
$\log g = 4.5$. The effect of $\log g$ on the infrared spectrum
is again rather modest and accurate determination of $\log g$
is difficult. 

\begin{figure*}
  \begin{center}
   \resizebox{0.9\hsize}{!}{
       \includegraphics{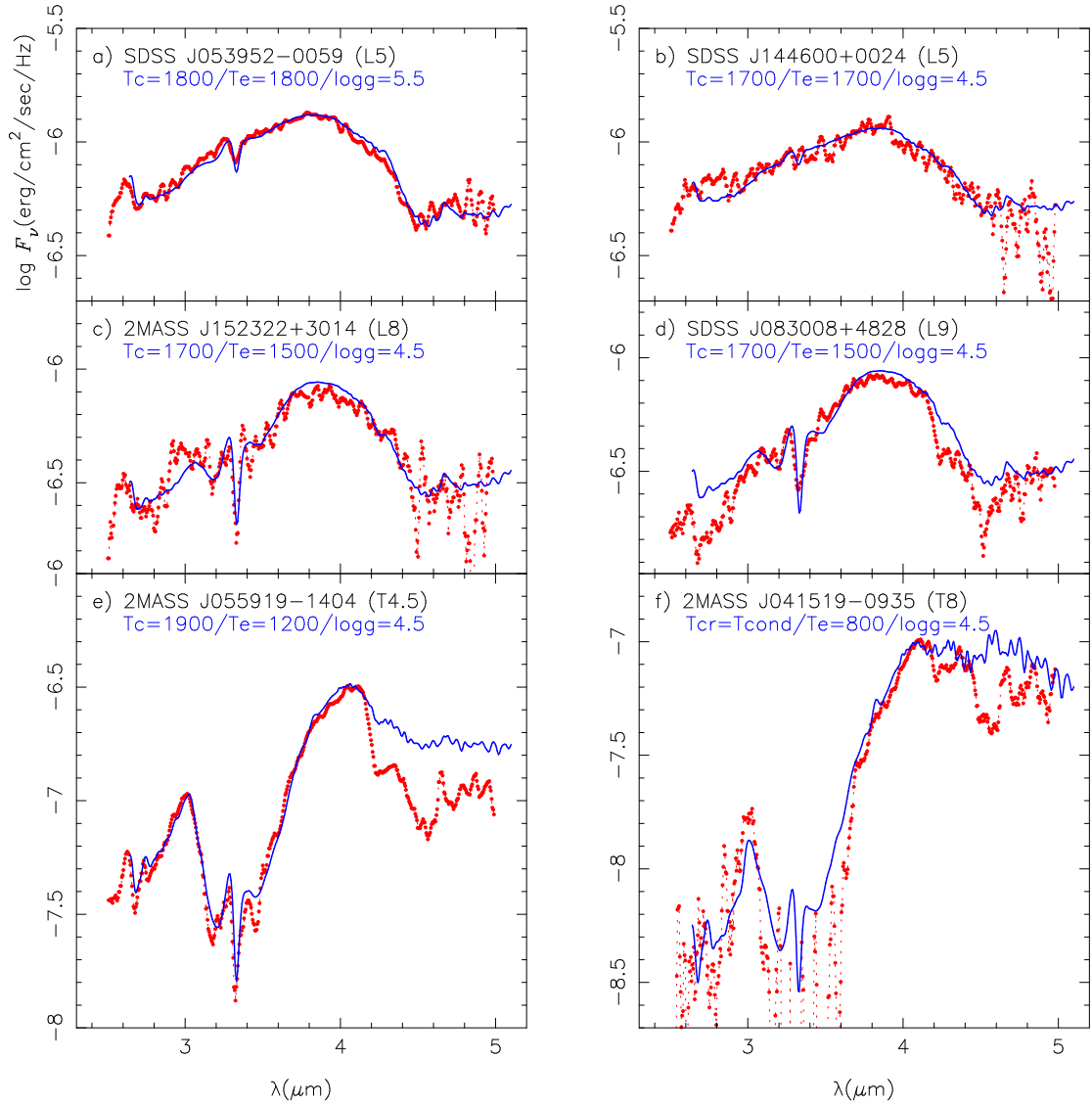}
   }
  \end{center}
\caption{ The observed and predicted spectra based on the UCMs are compared on
an absolute scale: a) SDSS J053951$-$0059 (L5); b) SDSS J144600$+$0024 (L5);  
c) 2MASS J152322$+$3014 (L8); d) SDSS J083008$+$4828 (L9); 
e) 2MASS J055919$-$1404 (T4.5); and f) 2MASS J041519$-$0935 (T8).
}\label{fig:fitall}
\end{figure*}

A detection of CO$_{2}$ at 4.2~$\mu$m suggested in Section~\ref{sec:nirspec} 
can be well supported by our model 
prediction that shows a depression at the expected position of 
the CO$_{2}$ 4.2~$\mu$m band and even shows a reasonable fit
quantitatively (see Figure~\ref{fig:fitall}c).
The fit in the region of the CO fundamental band 
appears to be poor and the deep absorption features near 4.5 and
4.9~$\mu$m remain unexplained. We suspect, however, that the CO
fundamental band provides the major contribution as in later type
dwarfs to be discussed in  Section~\ref{sec:fit0830}.

Thus the spectrum observed by \textit{AKARI} now shows some 
dissonances with the predicted spectrum in this late-L dwarf as shown 
in Figure~\ref{fig:fitall}c.

\subsubsection{SDSS J083008$+$4828}\label{sec:fit0830}
As already discussed in some detail in Section~\ref{sec:obsvsucm},
we find ($T_\mathrm{eff}$, $T_\mathrm{cr}$, $\log g$) = (1500, 1700, 4.5) 
for this object. The final fit is shown in Figure~\ref{fig:fitall}d. We 
notice that the resulting parameters are the same as that for 2MASS 
J152322$+$3014 discussed in Section~\ref{sec:fit1523}.
This result is also consistent with the similarity of the empirical 
$T_\mathrm{eff}$ and of the $J-K$ color between these two late-L dwarfs
(see Table~\ref{tbl:knowndata}). Nevertheless, the observed spectra
of the two objects differ significantly. For example,  the observed
feature  near the 2.7--2.9~$\mu$m region (largely due to H$_2$O $\nu_1, 
\nu_3$ and 2$\nu_2$ bands) shows a large difference between
these two objects of the same physical parameters.
Since this feature is well reproduced by our models in other objects except for
SDSS J083008$+$4828 (see Figure~\ref{fig:fitall}), this
should not be due to any systematic effects in our models.

A more interesting result is that the observed CO$_{2}$ 4.2~$\mu$m 
feature in this object is very strong as noted already 
(Section~\ref{sec:nirspec}), and that it is too strong compared with 
the prediction in marked contrast to the case of 2MASS J152322$+$3014. 
It appears that the CO$_{2}$ 4.2~$\mu$m feature cannot be 
explained by any combination of $T_\mathrm{eff}$, $T_\mathrm{cr}$, and 
$\log g$ with our UCM essentially based on the LTE assumption, while 
those of 2MASS J152322$+$3014 may be more consistent with the LTE prediction. 
Such a large difference in CO$_{2}$ band in the two objects of similar 
physical parameters cannot be explained by differences in dust cloud
properties if, as we find, $T_\mathrm{cr}$ is nearly the same around 1700K 
in both objects. The deep absorption at 4.5~$\mu$m can 
be identified with the $R$-branch of the CO fundamental, 
but cannot be explained within the framework of the LTE assumption as well.

Thus, there are serious difficulties in understanding the
spectrum observed by \textit{AKARI} in 
that  the region between 4 and 5~$\mu$m cannot be 
fitted at all by the LTE prediction based on our UCM.

\subsubsection{2MASS J055919$-$1404}\label{sec:fit0559}
Details of the model fitting of this object were described
in Section~\ref{sec:obsvsucm}, 
and our final fit on an absolute scale is shown in Figure~\ref{fig:fitall}e.
We are shocked to see that the discrepancy between the observed and
predicted spectra is so large in the region beyond 4.1~$\mu$m. 
The region of CO $R$-branch is too deep to be explained by our 
LTE model and CO $P$-branch also appears to be appreciable.  
Thus we confirm that the CO bands are actually enhanced not only in late-T 
dwarfs but also in mid-T dwarfs, as anticipated by photometric 
observations \citep[e.g.][]{Leggett07}.

In this object, a more important result is that  the new spectrum observed 
by \textit{AKARI} provides definite evidence for the presence
of the strong CO$_2$ band in brown dwarfs for the
first time. The depression at 4.2~$\mu$m is too large to be
accounted for by the predicted CO$_2$ band and the discrepancy
between the observed and predicted spectra is quite serious.
All these new features due to CO and CO$_2$
cannot be accounted for by the present LTE model such as our UCM.   
In contract to these serious disagreements in the region beyond
4~$\mu$m, the observed and predicted spectra agree quite well
in the region between 2.5 and 4.1~$\mu$m. 

Recent analysis of the SED of this object by 
\citet{Cushing08} concluded that $T_\mathrm{eff} =$ 1200\,K, $\log g =
5.5$  and $f_\mathrm{sed}=4$, 
where $f_\mathrm{sed}$ is the cloud sedimentation efficiency parameter
and a larger value implies greater sedimentation efficiency resulting
in a thinner cloud \citep{Ackerman01}. Thus $f_\mathrm{sed}$ is a 
parameter that may have a similar role as our $T_\mathrm{cr}$, with larger 
$f_\mathrm{sed}$ corresponding to higher $T_\mathrm{cr}$. Thus their result 
showing relatively large value of $f_\mathrm{sed} =$ 4 is consistent with 
our relatively high value of $T_\mathrm{cr} =$ 1900\,K, and we agree with 
them on the properties of the dust cloud in this T4.5 dwarf. We also agree 
that $T_\mathrm{eff}$ = 1200\,K for this object. Thus our results for the
cloud properties as well as for $T_\mathrm{eff}$ show good agreements with
those by \citet{Cushing08}, despite the fact that different
spectral regions are analyzed with different models. 
Unfortunately, the results for gravity differ considerably. However,
SEDs are not very sensitive to the gravity and this is the
parameter most difficult to determine from the model fittings
(as will further be discussed in Section~\ref{sec:gravity}).

In summary, some dissonances between the observed and predicted spectra,
seen in the late-L dwarfs, appear to be quite distinct
in this mid-T dwarf.

\subsubsection{2MASS J041519$-$0935}\label{sec:fit0415}
The empirical $T_\mathrm{eff}$ of this T8 dwarf is 764\,K 
(Table~\ref{tbl:knowndata}) and, 
if a dust cloud forms, 
it is situated below the layer of the optical depth 
unity in such a very cool dwarf. For this reason, the spectrum no
longer depends on $T_\mathrm{cr}$, as we have actually confirmed.
Accordingly, we apply the dust-free models in which dust grains have
all precipitated just after they are formed (this means $T_\mathrm{cr}$ 
= $T_\mathrm{cond}$), and we find the best possible fit for ($T_\mathrm{eff}$, 
$T_\mathrm{cr}$, $\log g$) = (800, $T_\mathrm{cond}$, 4.5). 

Determination of the physical properties of the coolest brown dwarfs
including 2MASS J041519$-$0935 has been discussed by \citet{Burgasser06a}, 
who calibrated the strengths of H$_{2}$O bands and 
relative fluxes in terms of $T_\mathrm{eff}$ and $\log g$. As a result,
they suggested $T_\mathrm{eff} = $740--760\,K and $\log g = $4.9--5.0
for this object. Also, \citet{Saumon07} suggested a similar
result of $T_\mathrm{eff} = $725--775\,K and $\log g = $5.00--5.37
based on the analysis of a spectrum including new data
obtained with \textit{Spitzer Space Telescope}. Our result 
is reasonably consistent with these results, although the values of
$\log g$ differ from each other. Thus analyses of the different spectral
regions by the different models show more or less consistent results
in this object at least for $T_\mathrm{eff}$.

The overall fit of the predicted spectrum of the possible best model 
is not so bad (see Figure~\ref{fig:fitall}f), but we confirm
that the region of the CO bands 
cannot be fitted at all. Such a possible large discrepancy in the
predicted and observed
CO bands in late-T dwarfs was previously suggested based on the ground 
based observation of Gl~229B \citep{Noll97, Oppenheimer98},
and more recently for two other  late-T dwarfs by \citet{Geballe09}. 
We confirm these previous results and add a new case of largely
enhanced CO fundamentals in late-T dwarfs. We also note
that the $R$-branch is much stronger compared with the $P$-branch. 

A more interesting result is that the CO$_{2}$ band at 4.2~$\mu$m 
appears quite distinctly in this late-T dwarf too. It is impossible to 
explain both the CO and also the CO$_{2}$ bands by
the LTE models, and we will discuss such cases in the 
Section~\ref{sec:nonlte}.

In conclusion, the new spectral region explored by \textit{AKARI} for the
first time reveals: first, the enhancement of the CO fundamental bands
is a general feature and, second, the CO$_{2}$ band at 4.2~$\mu$m appear to be 
an important absorber, in late-L and T dwarfs. 

\subsection{Basic physical parameters of six brown dwarfs}\label{sec:params}
We now discuss briefly the resulting $T_\mathrm{eff}$, $T_\mathrm{cr}$, 
$\log g$, and $R/R_\mathrm{J}$ values summarized in Table~\ref{tbl:fit}.
Also, we discuss briefly the effects of other basic parameters, namely
chemical composition (in Section~\ref{sec:cc}) and micro-turbulent velocity
(in Section~\ref{sec:vmic}), which we assumed to be the Solar.
If we try to determine these two parameters by the same way
as we have done in Section~\ref{sec:obsvsucm} and \ref{sec:ucmfitresults} 
for the three basic parameters, 
we must consider at least $3 \times 3 \times 3 \times 3 \times 3 = 243$ 
models instead of 27 models for each object,
but this is clearly impractical. Moreover, we do not think that the
chemical abundances and micro-turbulent velocity can be determined from
such low resolution spectra as those we have at hand. 

\subsubsection{Effective temperature and radius}\label{sec:teffr}
We notice first that our $T_\mathrm{eff}$'s do not agree with the
so-called empirical effective temperatures obtained from the observed 
luminosities and a constant radius such as $R = 0.9\,R_\mathrm{J}$\citep{Vrba04} 
(see 7-th column of Table~\ref{tbl:fit}). 
In fact, we already notice through Section~\ref{sec:fit0539} and \ref{sec:fit0415} that the 
observed SEDs cannot be fitted well with the predicted ones 
assuming the empirical $T_\mathrm{eff}$ by \citet{Vrba04}.
The reliability of the empirical $T_\mathrm{eff}$ largely depends on
the accuracy of the values of the radius.
The constant radius of $R = 0.9\,R_\mathrm{J}$ was 
based on a statistical analysis by \citet{Burgasser01} of the evolutionary 
models of cool dwarfs by \citet{Burrows97}, but it is unknown if
it is appropriate to apply such a value to all the objects.  

It is known that the radii of cool substellar
objects are independent of mass to within 30 per cent for a broad range
of masses from 0.3 to 70\,$M_\mathrm{J}$ \citep{Burrows01}.
The resulting radii of all our six objects (Table~\ref{tbl:fit}) appear to be
consistent with the prediction of the evolutionary models
and confirm that the radii of ultracool dwarfs are within approximately 
30 per cent of Jupiter's radius. The mean radius for our
six objects is 0.81\,$R_\mathrm{J}$, and thus we agree with
\citet{Burgasser01} in that the radii of
ultracool dwarfs are statistically smaller than Jupiter's radius. However, 
the variation of radii among brown dwarfs appears to be appreciable 
(Table~\ref{tbl:fit}) and may not be represented by a single mean value. 
For this reason, the so-called empirical effective temperature cannot be 
completely reliable. Although we use it as a guideline in our analysis 
(Sections~\ref{sec:obsvsucm} \&
\ref{sec:ucmfitresults}), it may even be misleading as we showed in the 
case of 2MASS J055919$-$1404 (Section~\ref{sec:obsvsucm}). 
 
Also, because of such variations of the radii in real ultracool dwarfs,
an attempt to analyze the observed spectra on an absolute scale 
based on an assumed radius of $R = R_\mathrm{J}$ \citep{Tsuji05} does not 
necessarily work well.
For example, we showed before that  the near infrared spectrum  of
2MASS J152322$+$3014 observed with Subaru, reduced to an absolute scale with 
the known parallax and with the assumption of $R = R_\mathrm{J}$, agreed  
with the predicted spectrum of $T_\mathrm{eff} =$ 1300\,K.  For this
reason, we suggested $T_\mathrm{eff}$
of this object to be $\approx 1300$\,K \citep{Tsuji05}.
However, we find that the shape of the \textit{AKARI} spectrum can be best
explained by
 $T_\mathrm{eff} =$ 1500\,K (Section~\ref{sec:fit1523}), and 
that a consistency with the SED on an absolute scale can be 
achieved if we abandon the ad hoc assumption of $R = R_\mathrm{J}$ 
(see Figure~\ref{fig:fitall}c).

In conclusion, we believe that the observed spectrum (or relative SED)
can provide a more direct and realistic estimation of the effective 
temperature of an individual object. Further, an empirical estimation of
the radius can be possible if the SED can be reduced to an absolute
scale with the known parallax. With this conclusion, we now think it
inappropriate to analyze the observed spectrum in absolute scale
based on an ad hoc assumption of $R = R_J$, as we proposed before
\citep{Tsuji05}.

\subsubsection{Gravity}\label{sec:gravity}
The determination of $\log g$ is more difficult since the low-resolution infrared
spectra depend little on $\log g$, as can be seen 
in our Figures~\ref{fig:fit0830} \& \ref{fig:fit0559}.
Also, this difficulty is clearly shown by a goodness-of-fit 
(statistic $G_k$ defined by Equation~(1) in \citet{Cushing08} and shown
in their Figure~5, in which $G_k$ only shows minor changes for different
$\log g$'s (possibly except for no cloud case). 

Inspection of Table~\ref{tbl:fit} reveals that 
our result appears to be biased towards low
gravity, namely $\log g = 4.5$ for 5 out of 6 objects. This may be due to
the difficulty of gravity determination noted above, at least partly,
or may be simply  because our sample mostly consists of low gravity objects 
by chance. We examine if this may be due to any systematic effects in our
UCMs. Our gravity determination is essentially based on the gravity 
dependence of the gas and dust opacities, but
we are using more or less standard methods in treating molecular
and dust opacities. Also, a
comparative study of different modeling approaches did not reveal
any systematic effect in our models \citep{Helling08}.
Thus we cannot identify any systematic effect in our models. We
hope to examine this problem further with a larger sample in future. 

\citet{Cushing08} used ultracool dwarf evolution
models to transfer their ($T_\mathrm{eff}, R$) values to
($T_\mathrm{eff}, g$) values, from which they obtained $\log g = 4.7$
for 2MASS J055919$-$1404. This result differs from their result based on 
their spectral fitting referred to in Section~\ref{sec:obsvsucm} ($\log g =5.5$),
but agrees rather well with our value given in Table~\ref{tbl:fit} 
($\log g = 4.5$). 
Thus this result can be supporting evidence for our $\log g$, despite
some difficulties noted above. However, the results based on the evolutionary
sequences may suffer more or less similar difficulty to the spectral
fitting method, since it depends critically on the physical parameters
such as $T_\mathrm{eff}$, which cannot be very accurate if they are
 based on the spectral fitting in the limited
spectral region (as will be detailed in Section~\ref{sec:errors}).

Given the radius and $\log g$, it is in principle possible to estimate
the mass. However, SEDs are not so sensitive to $\log g$ and its accuracy is
rather low. For this reason, we believe it impossible to 
obtain a reasonable mass estimates from our spectra.

\subsubsection{Critical temperature}\label{sec:tcr}
Inspection of Table~\ref{tbl:fit} reveals that the resulting 
$T_\mathrm{cr}$'s show
a variety of values in our limited sample, and this result
implies that the dust cloud properties differ significantly among
different objects. In modeling  dusty dwarfs,
it is still difficult to determine the dust cloud properties from the 
basic physics, and we had to introduce such an empirical  parameter as 
$T_\mathrm{cr}$ to represent the dust cloud properties, at least partly.
Other models of  dusty dwarfs also have a more or less similar feature. 
For example, \citet{Ackerman01} introduced the cloud sedimentation efficiency
$f_\mathrm{sed}$ in their models and, as discussed in Section~\ref{sec:fit0559}, 
their $f_\mathrm{sed}$ and our $T_\mathrm{cr}$ consistently showed that the dust 
cloud in 2MASS J055919$-$1404 is rather thin. We hope that a larger
sample will be analyzed to clarify the nature of dust clouds and that the 
result will serve as a guide towards a more physical theory of cloud 
formation in dusty dwarfs. 

\subsubsection{Chemical composition}\label{sec:cc}

We assumed that the chemical composition in such unevolved objects as
brown dwarfs should be the same as the Solar composition. However,
the Solar composition itself is by no means well established yet. 
For example, we adopted $\log A_\mathrm{C} = 8.60$ and $\log A_\mathrm{O} = 8.92$ 
(on the scale of
$\log A_\mathrm{H} = 12.0$) in our initial version of UCM based on the latest 
results known at that time \citep{Tsuji02}. However, these values were
revised to be $\log A_\mathrm{C} = 8.39$ and $\log A_\mathrm{O} = 8.69$  based on 
the three-dimensional(3D) time-dependent  hydrodynamical model of the 
solar photosphere \citep{AllendePrieto02}
and our present version of UCM is based on these revised values
as noted previously \citep{Tsuji04}. These values are close
to the more recent  values of $\log A_\mathrm{C} = 8.43$ and 
$\log A_\mathrm{O} = 8.69$ \citep{Asplund09}.
Since a direct determination of abundances in brown dwarfs cannot
be expected in the near future, we think it best for now to use the
most reliable Solar composition.

Unlike the cases of evolved stars in which the surface chemical 
composition may suffer drastic variations due to convective dredge-up of
the products of nuclear processing in the interior, large variations of
the surface chemical composition may not be expected in brown dwarfs, 
but a possibility of small variations may  not be excluded. 
As an example, we examine the effect of carbon abundance on CO and CO$_2$ 
features in J083008$+$4828 and J055919$-$1404, since CO and  CO$_2$ features in these objects 
could not be  reproduced well with the composition  we have assumed. 
For this purpose, we compute two spectra with $\log A_\mathrm{C}$ changed by 
$\pm 0.15$\,dex compared to the standard 
composition we have assumed, namely, $\log A_\mathrm{C} = 8.39$. 

\begin{figure*}
  \begin{center}
   \resizebox{0.6\hsize}{!}{
       \includegraphics{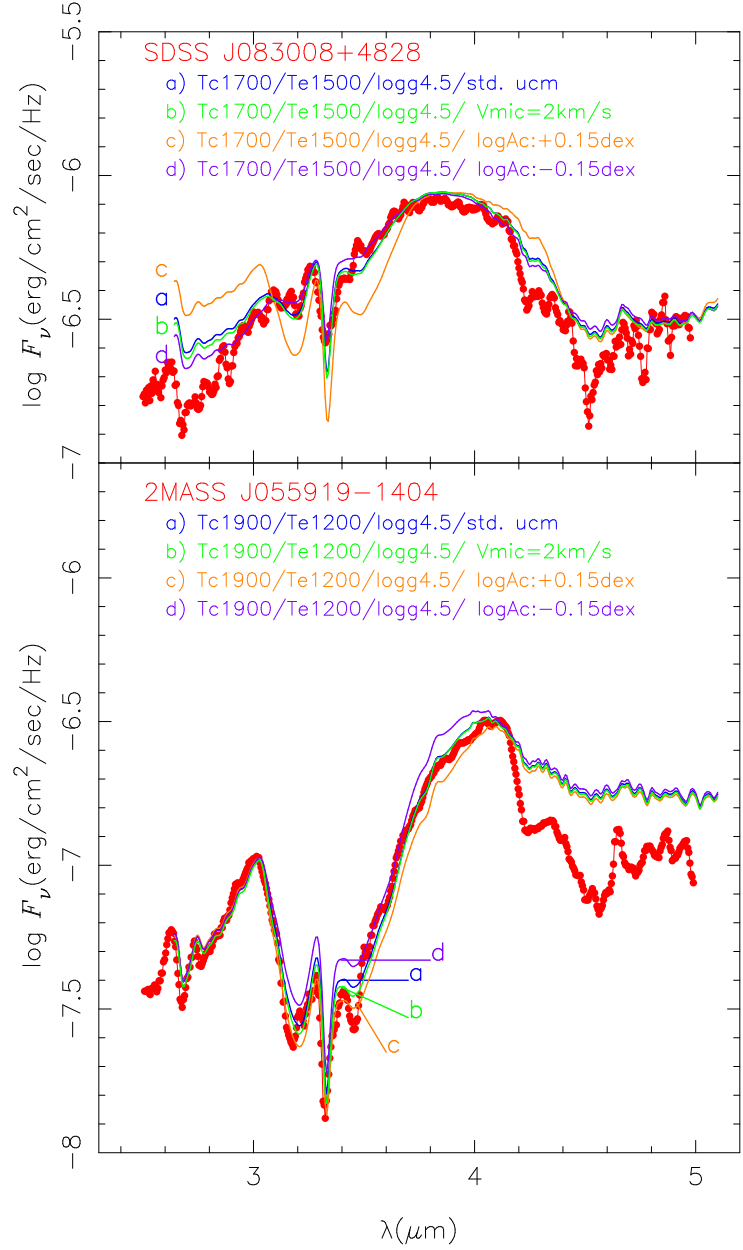}
   }
  \end{center}
\caption{The effects of changing $\log A_\mathrm{C}$ by $\pm 0.15$\,dex
 and of increasing
the micro-turbulent velocity from 1 to 2 km\,s$^{-1}$ are shown
for J083008$+$4828 in the upper diagram. The same effects are shown for
J055919$-$1404 in the lower diagram.
}\label{fig:ccvmic}
\end{figure*}

The results for J083008$+$4828 are shown in the upper diagram of Figure~\ref{fig:ccvmic}. 
The effect of increasing $\log A_\mathrm{C}$ by 0.15\,dex results in a considerable 
strengthening of the methane bands, but the overall fitting tends
to be worse (compare c with a in the figure). The CO$_2$ and CO
bands at 4.2 and 4.6 $\mu$m, respectively, however, do not show
any strengthening. The effect of decreasing $\log A_\mathrm{C}$ 
by 0.15\,dex is rather modest (see d in the figure).
The results for J055919$-$1404 are shown in the lower diagram of Figure~\ref{fig:ccvmic}. 
The effect of changing $\log A_\mathrm{C}$  by $\pm 0.15$\,dex
can be noticed on the CH$_4$ bands, but CO$_2$ and CO bands remain
almost unchanged. We conclude that the large discrepancies between
the observed and predicted CO and CO$_2$  bands in J083008$+$4828 and J055919$-$1404
cannot be the problem of the carbon abundance.

\subsubsection{ Micro-turbulent velocity}\label{sec:vmic}

We have assumed the micro-turbulent velocity to be the Solar value
of 1 km\,s$^{-1}$ throughout. Just to see the effect of the
micro-turbulent velocity, we increase it to be 2 km\,s$^{-1}$ and
the resulting spectra are compared with those for the case of
1 km\,s$^{-1}$ for J083008$+$4828 and J055919$-$1404 in the upper and lower
diagrams, respectively, of Figure~\ref{fig:ccvmic}. Clearly, dependence on the 
micro-turbulent velocity is quite small (compare a and b
in Figure~\ref{fig:ccvmic}) as expected for such
high density photospheres dominated by the pressure broadening.

\subsection{Errors and limitations of the present  analysis based on the UCMs}\label{sec:errors}

Since we examine the effect of $T_\mathrm{eff}$ by steps of 100\,K, 
our $T_\mathrm{eff}$ cannot be more accurate than $\pm50$\,K. 
Further, combined with the similar uncertainty in $T_\mathrm{cr}$, 
the final uncertainty in our estimation of $T_\mathrm{eff}$ may be
$\pm 100$\,K. But we should notice that the 
parameters we have found are simply those that explain the \textit{AKARI} 
spectra in the 2.5--5.0~$\mu$m region. We tried more or less 
similar analysis of the 1.0--2.5~$\mu$m region of our objects and found some
differences in the resulting parameters, although the differences
are mostly within the errors outlined above. In other words, the solution
obtained by fitting other spectral regions may not in general explain the
2.5--5.0~$\mu$m spectra.
As summarized in Figure~\ref{fig:fitall}, we are trying 
to reach a unified understanding
of the series of new spectra observed for the first time by \textit{AKARI} 
covering from mid-L to late-T dwarfs. For this purpose,
we believe it best to apply the parameters obtained from the
\textit{AKARI} spectra themselves and analyze all our sample consistently.

As shown by \citet{Cushing08}, the effective temperatures obtained by 
fitting a limited spectral region differ from those obtained by fitting 
the full SED (0.95--14.5~$\mu$m) by typically $\approx$ 200\,K and by 
as large as 700\,K in the worst case. As noted above, we 
confirm essentially the same difficulty with our UCM. The reason for
this difficulty may be because the present models are far from
perfect. For example, we do not know the exact form of the dust opacities
which suffer from many unknown effects (e.g. composition, size, shape,
impurity etc.) and of the molecular opacities based on imperfect
line lists. Moreover, we do not know yet the exact nature of the dust
clouds formed in the photosphere of cool dwarfs. 
For these reasons, we think it difficult  to expect such rigorous numerical 
accuracy for brown dwarfs as realized in model photospheres of 
ordinary stars. 
We intended from the beginning that our 
models can at least be of some help as a guide in interpretation and 
analysis of the observed data of ultracool dwarfs \citep[e.g.][]{Tsuji01}.
  
A more perfect model would allow us to more safely analyze a wider
spectral region. However, current models of brown dwarf atmospheres
are far from perfection. Fitting of a wider spectral region may sample 
more regions of uncertain opacities at the same time, and is by no 
means easier to have better results.
Thus we must allow for additional uncertainties beyond those
found by fitting one spectral region.
We fully agree 
with \citet{Cushing08} who concluded that the accurate determination of 
the physical parameters is  still an elusive goal, even though SED fitting 
method shows reasonable success in general. 

For now, we must be satisfied in that our SED fitting  analysis based on 
our UCM has also been done fairly consistently for the new spectral region 
observed by \textit{AKARI}.
The UCM is a semi-empirical model in which the property of the dust 
clouds is represented by a single parameter $T_\mathrm{cr}$ and, for 
this very reason, our UCM is free from yet unknown details of the 
cloud formation processes. Thus, our UCM can be flexible enough to 
interpret the major characteristics of the observed spectra in terms of 
a few basic physical parameters. The UCM, however, assumes LTE throughout 
and hence cannot be applied directly to the observed features showing possible 
deviations from LTE. We will discuss such a case in the next section.

\begin{figure}[!ht]
  \begin{center}
   \resizebox{0.9\hsize}{!}{
       \includegraphics{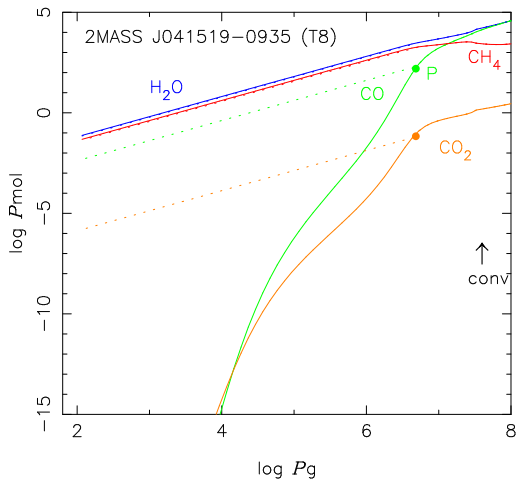}
   }
  \end{center}
\caption{Molecular abundances, presented as partial pressure
vs total pressure (pressure in units of dyn\,cm$^{-2}$), in LTE (solid
 lines) and in non-equilibrium modified by vertical mixing for the case
of $b=0.1$ (dashed lines)
in the photosphere of the T8 dwarf 2MASS J041519$-$0935 
(Model: $T_\mathrm{eff}$/$T_\mathrm{cr}$/
$\log g = $800\,K/$T_\mathrm{cond}$/4.5). The arrow
indicates the onset of convection.
}\label{fig:nlte0415}
\end{figure}

\begin{figure}[!ht]
  \begin{center}
   \resizebox{0.9\hsize}{!}{
       \includegraphics{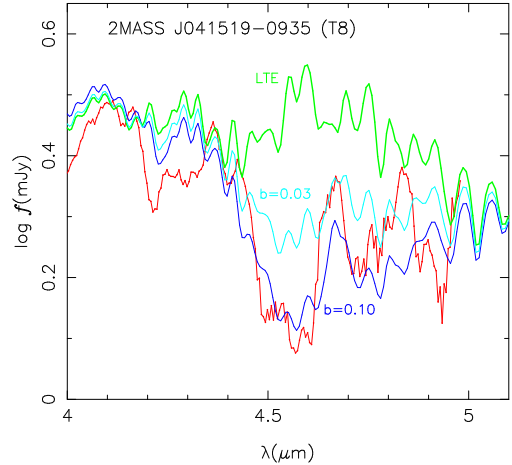}
   }
  \end{center}
\caption{ 
The effect of vertical mixing on the CO spectrum in
2MASS J041519$-$0935 (T8): The cases of LTE, $b = 0.03$, and $b = 0.10$ 
are compared with the observed spectrum.
}\label{fig:fitco0415}
\end{figure}

\section{Spectral features unexplained by the LTE models}\label{sec:nonlte}

In the previous section, the overall SEDs can be understood reasonably well 
with the UCM based on LTE, but the observed features of CO and CO$_2$ in 
T-dwarfs (and possibly late-L dwarfs) cannot be fitted at all with 
the spectra predicted by the model. As for CO, this fact has been 
known for late-T dwarfs \citep[e.g.][]{Noll97, Oppenheimer98, Geballe09}, 
and we confirm this fact by the better quality data for 
T8 dwarf 2MASS J041519$-$0935 (Figure~\ref{fig:fitall}f).
But we also find that CO bands are enhanced significantly
already in the mid-T dwarf 2MASS J055919$-$1404
(Figure~\ref{fig:fitall}e) as well as 
in the late-L dwarf SDSS J083008$+$4828 (Figure~\ref{fig:fitall}d).
A new and more difficult problem is that the CO$_2$
band at 4.2~$\mu$m is enhanced enormously in all the cool dwarfs later than L9.

\subsection{Late-T dwarf}\label{sec:nonlte_lateT}

The unexpected detection of CO in the late-T dwarf Gl~229B
has been interpreted as
due to vertical mixing of CO from the warm layers where CO is
still abundant to the cooler layers where CO is mostly transformed
to CH$_4$ under LTE. This is possible since CO is so stable that the 
time scale $\tau_\mathrm{chem}$ of the reaction
 $$ \mathrm{CO} + 3\mathrm{H}_2  \rightleftharpoons
        \mathrm{CH}_4 + \mathrm{H}_2\mathrm{O}                  $$ 
is very slow compared to the mixing time scale $\tau_\mathrm{mix}$ 
\citep{Griffith99, Saumon00}, and a
detailed computation including such a non-equilibrium effect has 
been carried out \citep[e.g.][]{Saumon07, Hubeny07}. 

We try to explain our observations by a simple computation as follows: 
We show the LTE molecular abundances of CO, H$_2$O, CH$_4$, and CO$_2$ in 
the photosphere of 2MASS J041519$-$0935, 
using the UCM of ($T_\mathrm{eff}$, $T_\mathrm{cr}$, $\log g$)
 = (800\,K, $T_\mathrm{cond}$, 4.5), by the solid lines in 
Figure~\ref{fig:nlte0415}. We assume that the time scale of
CO destruction $\tau_\mathrm{chem}$ is equal to the
timescale of mixing $\tau_\mathrm{mix}$ at a depth point P in 
Figure~\ref{fig:nlte0415}
and CO/CH$_4$ abundance ratio is fixed at the value $b$ of that
point. We examine several values of $b$. The case of
$b = 0.10 $ is shown in Figure~\ref{fig:nlte0415} as an example. The resulting
non-equilibrium abundance of CO is indicated by the dashed line. If H$_2$O 
and CO$_2$ abundances relative to CO are fixed at the values at P as are 
CH$_4$, H$_2$O as well as CH$_4$ abundances in the upper layers should be 
decreased and the results are also shown by the dashed lines. The decreases,
however, are rather minor and difficult to see in the figure. On the 
other hand, CO$_2$ abundance shows a large increase as seen by 
the dashed line in Figure~\ref{fig:nlte0415}. 

We compute the spectra based on the non-equilibrium abundances 
for $b = 0.03 $ and $b = 0.1$ together with that of the full LTE, 
and the results are 
compared with the observed spectrum of 2MASS J041519$-$0935 in 
Figure~\ref{fig:fitco0415}. Inspection of the figure reveals 
that the observed CO spectrum of 2MASS J041519$-$0935 can roughly 
be accounted for with $b = 0.1$. However, the predicted CO$_2$ band is 
still too weak compared with the observation even for $b = 0.1$.
Of course, there is no reason why CO$_2$/CH$_4$ ratio can be 
fixed at the value at point P, since chemical reaction time-scale of CO$_2$
cannot be the same as for CO. Further, there is little possibility
that more CO$_2$ can be dredged-up, since CO$_2$ abundance in the
deeper layers is not as large as shown in Figure~\ref{fig:nlte0415}. 
 
On the other hand, it is possible that CO$_2$ abundance
will change to the equilibrium value of 
the local physical condition
if the chemical time scale related to CO$_2$ formation is not as slow
as in the case of CO.
However, if CO$_2$ attains its equilibrium value with the abundant
CO resulting from vertical mixing, CO$_2$ abundance will be too
large. In fact, CO$_2$ is more abundant than CO at $\log P_\mathrm{g} < 4.0$ 
in LTE as can be inferred from the solid lines in 
Figure~\ref{fig:nlte0415}. 
For this reason, we cannot assume simply that CO$_2$ can be in
equilibrium with CO. We cannot predict the precise value of CO$_2$ 
abundance without more detailed non-equilibrium analysis, but we can 
expect from the above consideration that
the CO$_2$ abundance can be somewhat larger than those shown by
the dashed line in Figure~\ref{fig:nlte0415} in the surface layers of 
2MASS J041519$-$0935 and hence the CO$_2$ band could then be  stronger 
than those suggested in Figure~\ref{fig:fitco0415}.

\subsection{Mid-T dwarf}\label{sec:nonlte_midT}
\begin{figure}[!ht]
  \begin{center}
   \resizebox{0.9\hsize}{!}{
       \includegraphics{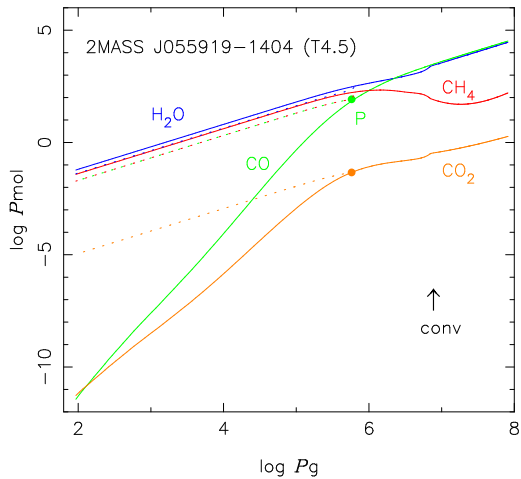}
   }
  \end{center}
\caption{Molecular abundances in LTE (solid lines) and
in non-equilibrium by vertical mixing for the case
of $b=0.5$ (dashed lines)
in the photosphere of T4.5 dwarf 2MASS J055919$-$1404 
(Model: ($T_\mathrm{eff}$, $T_\mathrm{cr}$, $\log g$) = (1200\,K, 1900\,K4.5)). 
The arrow indicates the onset of convection.
}\label{fig:nlte0559}
\end{figure}

A new result of the \textit{AKARI} observation is that the  
mid-T dwarf 2MASS J055919$-$1404 also shows much stronger 
CO and CO$_2$ bands than those 
predicted with the use of the UCM based on LTE. 
We show the LTE molecular 
abundances in the photosphere of 2MASS J055919$-$1404, using the UCM of 
($T_\mathrm{eff}$, $T_\mathrm{cr}$, $\log g$) = (1200\,K, 1900\,K, 4.5), 
by the solid lines in Figure~\ref{fig:nlte0559}.

We first try the vertical mixing model used for the late-T dwarf
2MASS J041519$-$0935. We again assume that the time scale of
CO destruction $\tau_\mathrm{chem}$ is equal to the timescale of mixing 
$\tau_\mathrm{mix}$ at a point P in Figure~\ref{fig:nlte0559}
and the CO/CH$_4$ abundance ratio is fixed to a value $b$ at this
point. We again test a few values of $b$. A case of
$b = 0.50 $ is shown in Figure~\ref{fig:nlte0559}. The resulting
non-equilibrium abundance of CO is indicated by the dashed line.
Also, H$_2$O and CO$_2$ abundances relative to CO are fixed at 
the values at P and shown by the dashed lines in Figure~\ref{fig:nlte0559}. 

\begin{figure}[!ht]
  \begin{center}
   \resizebox{0.9\hsize}{!}{
       \includegraphics{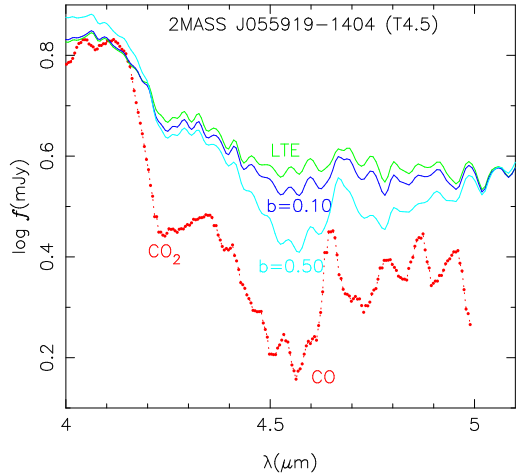}
   }
  \end{center}
\caption{Predicted spectra in LTE and in non-equilibrium by
vertical mixing ($b=0.10$ and $0.50$) compared with the
observed one of T4.5 dwarf 2MASS J055919$-$1404.
}\label{fig:fitco0559}
\end{figure}

We compute the spectra based on the non-equilibrium abundances 
for $b = 0.1$ and $b = 0.5$ 
together with those based on the LTE values, and the results are 
compared with the observed spectrum of 2MASS J055919$-$1404 in 
Figure~\ref{fig:fitco0559}. 
The predicted spectra for non-equilibrium
CO and CO$_2$ abundances show additional depression in the region of
CO $R$+$P$-branches but little change in the 4.2~$\mu$m region,
compared to the predicted one for the LTE case. 
This result indicates that the observed CO band can be explained 
at least partly by the vertical mixing model, but the observed
CO$_2$ remains completely unexplained.  The observed CO feature
is still too deep compared with the predicted one, but this may 
be due to the effect of  the deep CO$_2$ feature whose origin is
still unknown.

We conclude that the very strong CO  band observed
in 2MASS J055919$-$1404 by \textit{AKARI} can be at least partly explained by 
the vertical mixing model but the CO$_2$ feature cannot at all. 

\subsection{Late-L dwarf}\label{sec:nonlte_lateL}

We detected a deep absorption around 4.5~$\mu$m in the late-L dwarf 
SDSS J083008$+$4828 and it can be identified with the CO $R$-branch. 
The $R$-branch tends to be deeper than $P$-branch in CO
because of narrower line separation.
We notice such a structure of the CO fundamental consisting of the deep 
$R$-branch and weak $P$-branch in 2MASS J055919$-$1404, and this similarity 
lends support in identifying the 4.5~$\mu$m feature in SDSS J083008$+$4828 as 
due to the $R$-branch of CO in the relatively low-quality spectrum of 
this source (see also, Figure~\ref{fig:ircspec}), even though 
the presence of such a deep CO feature is quite unexpected in L-dwarfs.
We also see that the 4.2~$\mu$m feature is very strong.

\begin{figure}[!ht]
  \begin{center}
   \resizebox{0.9\hsize}{!}{
       \includegraphics{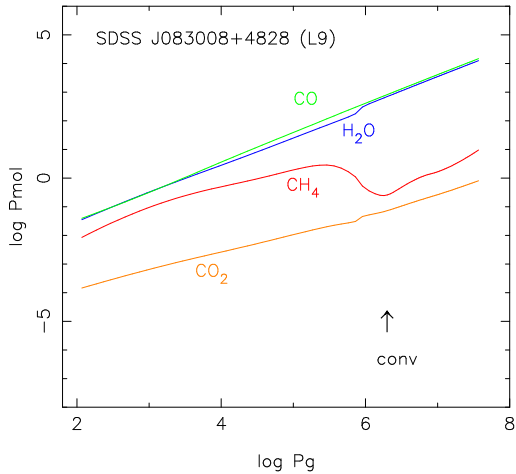}
   }
  \end{center}
\caption{Molecular abundances in LTE (solid lines) in the 
photosphere of the L9 dwarf SDSS J083008$+$4828
(Model: ($T_\mathrm{eff}$, $T_\mathrm{cr}$, $\log g$) = (1500\,K, 1700\,K, 4.5)). 
A sudden change of H$_2$O and CO$_2$
abundances at $\log P_\mathrm{g} \approx 6.0$ is due to formation
of silicate (MgSiO$_3$). The arrow indicates the onset of convection.
}\label{fig:nlte0830}
\end{figure}

In SDSS J083008$+$4828, the observed CO band is quite strong while CH$_4$ band
is fairly weak. These observed features imply that CO is quite abundant, and we
confirm this possibility in Figure~\ref{fig:nlte0830} where LTE abundances 
of CO, CO$_2$, CH$_4$, and H$_2$O are shown based on the UCM of 
($T_\mathrm{eff}$, $T_\mathrm{cr}$, $\log g$) = (1500\,K, 1700\,K, 4.5). 
Inspection of Figure~\ref{fig:nlte0830} reveals that the CO abundance under
LTE is already at its maximum possible value (i.e. almost all the carbon is 
in CO). Then, vertical mixing, if present, cannot supply additional 
CO to the upper photosphere and has little effect on the CO band
strengths. 
Thus, without any computation of the spectrum, it is clear 
that the vertical mixing model cannot be applied to this late-L dwarf. 
Given that an increase of the CO abundance in the photosphere can 
no longer be possible (Figure~\ref{fig:nlte0830}), not only by vertical 
mixing but also by any other method, some changes in the structures of the 
photosphere and/or in the atmosphere are required to produce 
the unusually deep CO band observed. 

The large strengthening of the CO$_2$ band cannot be due to vertical 
mixing either, since CO$_2$ abundance relative to other molecules 
is smaller in the deeper layers (Figure~\ref{fig:nlte0830}) and hence 
vertical mixing will
reduce rather than enhance the CO$_2$ abundance in the upper
layers. This result suggests that a non-equilibrium process other than 
vertical mixing is needed to explain the large enhancement 
of the CO$_2$ band.

\subsection{Vertical mixing and/or other possibilities}
Our analysis confirms that the vertical mixing model
can be applied to T-dwarfs at least for CO.
This result certainly shows that the vertical mixing model 
provides one possible way to explain the CO anomaly, but this
fact may not necessarily imply that it is a unique solution
for this phenomenon. However, our analysis cannot prove the 
applicability of the vertical mixing model to the  late-L dwarfs.

One problem is that  the convective zone is actually situated rather deep
in the photosphere of cool dwarfs \citep[e.g.][]{Tsuji02},
although it is said that cool dwarfs are fully convective.
For example, the upper boundary of the convective zone is
indicated by the arrow in
Figures~\ref{fig:nlte0415} \& \ref{fig:nlte0559},
and the layers where vertical mixing is expected are well
above the convective zone. Hence vertical mixing
cannot be due to convective mixing. Then, the problem is  
the unknown mechanism of efficient mixing in the photosphere of T-dwarfs.

A more difficult problem is how to understand the enhancement of CO$_2$ bands 
at 4.2~$\mu$m  in late-L and T dwarfs throughout. We confirm that this 
phenomenon cannot be explained by the vertical mixing model at all.
So the  question is whether the enhancements of CO and CO$_2$ bands
in late-L and T dwarfs are due to the same mechanism. It is still possible 
that an unknown process resulting in the enhancement of CO$_2$ will also 
result in the CO anomaly. In any case, the vertical mixing model 
cannot have general applicability, and we must
look for a more unified model that can be applied to CO and 
CO$_2$ in L and T-dwarfs. Presently, we have no solution yet
as to the reason why CO$_2$ (and also CO at least partly) are so strong 
in these objects. 
If non-equilibrium processes play major roles, we wonder why
the departure from LTE can be so significant in the brown dwarf
photosphere, where local density is so high that LTE condition
can be naturally satisfied.

More recently, \citet{Stephens09} analyzed the 0.8--14.5~$\mu$m region 
spectra of many L--T dwarfs and determined not only 
effective temperature, surface gravity and grain sedimentation efficiency  
but also vertical gas transport efficiency, based on the models of
\citet{Saumon08}. They showed that vertical mixing improves the fits 
and the observed spectra were fitted well in general. It will be interesting
to check whether  
their models can explain the enhanced CO bands in late-L dwarfs.  

\section{Discussion and Concluding Remarks}

\textit{AKARI} provides the first and unique opportunity
to take brown dwarf spectra in the important wavelength range 
2.5--5.0~$\mu$m continuously without interference of telluric atmosphere.
In particular, molecular bands of CO, CH$_4$, and CO$_2$
altogether in one spectrum present a great advantage for
investigating chemical processes in the atmospheres of 
ultracool dwarfs, through quantitative analysis of their carbon budgets.
\textit{AKARI} reveals many remarkable features of ultracool dwarfs.
Our observations confirm the presence of CO molecules
in the brown dwarf atmospheres of all spectral types
from L to very late T-type objects.
The fact that all observations ever made for the late-T
dwarfs detected CO molecules implies that the non-equilibrium
chemical composition is a common property of the late-T dwarfs.

It has been discussed that this non-equilibrium abundance 
is due to vertical mixing in the upper photosphere.
The CO fundamental band in the T8 dwarf 2MASS J041519$-$0935 
and the T4.5 dwarf 2MASS J055191$-$1404 can be
accounted for by the vertical mixing model (Section~\ref{sec:nonlte_lateT}).
The idea is also supported by \citet{Saumon06},
and by \citet{Mainzer07} based on the analysis of the mid-infrared
spectrum of T7.5 dwarf Gl~570D taken by \textit{Spitzer}/IRS.
They found that the abundance of the NH$_3$ molecule in the object is
almost one order of magnitude smaller than that expected under 
equilibrium conditions. It is considered that fresh N$_2$
molecules are continuously provided from inside.
Also, the vertical mixing model is shown to be consistent with 
the photometry of a larger sample of brown dwarfs obtained with 
\textit{Spitzer}/IRAC \citep{Patten06, Leggett07}.

However, our observations reveal that the CO fundamental band 
is stronger than predicted by the LTE models in the 
late-L dwarf (SDSS J083008$+$4828).
Since, unlike in T-dwarfs, CO is already
highly abundant in the upper photosphere of these objects,
no appreciable increase of CO abundance and hence of the CO
band strength can be expected by additional CO from the 
deeper warm layers. For this reason,
some mechanism(s) other than vertical mixing should be
looked for in late-L dwarfs. 
This quest is further strengthened by the detection of
CO$_2$ by \textit{AKARI} as discussed below. 

A new piece of information provided by \textit{AKARI} is the detection of CO$_2$
molecule. The CO$_2$ band is never observable from the ground,
and its 4.2~$\mu$m band is just outside of \textit{Spitzer}/IRS coverage.
Gas phase CO$_2$ in stellar atmospheres was first detected
in late-type giants by \textit{ISO}/SWS observations.
It was rather unexpected that in addition to the stretching-mode
absorption band at 4.2~$\mu$m, the bending mode ro-vibrational
bands in the 15~$\mu$m region are often seen in these stars,
sometimes in emission \citep{Ryde98, Justtanont98}.
The CO$_2$ band is generally stronger than those expected
from the LTE models, implying that some kinds of non-equilibrium
processes, either formation of the molecule or radiative excitation,
might play a role in the atmospheres of such stars.
The situation is of course quite different in brown dwarfs,
with lower temperature and much higher surface density than
in red giants.
Some amount of the CO$_2$ molecule is expected under LTE
conditions in brown dwarf atmospheres. Though the relative
abundance with respect to that of H$_2$O and CH$_4$ is small
(Figures~\ref{fig:nlte0415},~\ref{fig:nlte0559},~\&~\ref{fig:nlte0830}), 
the high $f$-value of this molecule helps to form 
a rather strong absorption band at 4.2~$\mu$m region as shown 
in Section~\ref{sec:predspec} (see Figure~\ref{fig:co2}). However, a simple
comparison with the model tells us that observed absorption
is much stronger than expected. 

\begin{figure}[!ht]
  \begin{center}
   \resizebox{1.0\hsize}{!}{
       \includegraphics[50,150][450,570]{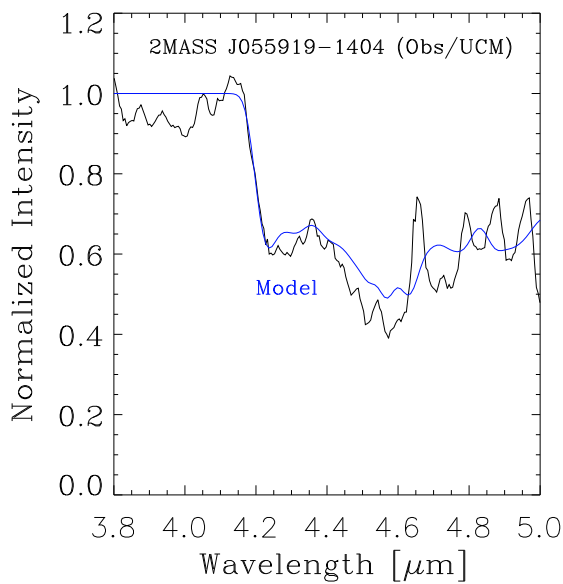}
   }
  \end{center}
  \caption{
The \textit{AKARI} spectrum of 2MASS J055919$-$1404 is divided by the
best fit UCM model and shown in black. Blue line indicates
the fitted model spectrum by a plane-parallel configuration
with the excitation temperature and the column density of
($T_\mathrm{ex}$, $N$)$_\mathrm{CO_2}$ = (800 K, $5\times 10^{17}$\,cm$^{-2}$) and
($T_\mathrm{ex}$, $N$)$_\mathrm{CO}$ = (1000 K, $2\times 10^{20}$\,cm$^{-2}$),
respectively.
}\label{fig:extfit}
\end{figure}

We estimate the temperatures and column densities of the extra CO$_2$
and CO absorption components in 2MASS J055919$-$1404
and SDSS J083008$+$4828. The observed spectra are divided by the
best fit UCM spectra shown in Figure~\ref{fig:fitall}, and the residual
spectra are fitted by two-layers plane-parallel model
with given excitation temperature $T_\mathrm{ex}$ and column density $N$
for each molecule.
In this simple model we adopt a Gaussian line width of FWHM = 30~km\,s$^{-1}$ based upon
measurements of H$_2$O lines by high-resolution spectroscopy 
in the J-band \citep{McLean07}. 
Because of the simplicity of the model, the fitting results
are suitable mainly for order of magnitude discussions.
The excitation temperature is only mildly constrained by the band shape.
Nevertheless, we obtain
($T_\mathrm{ex}$, $N$)$_\mathrm{CO_2}$ = (800 K, $5\times 10^{17}$\,cm$^{-2}$) and
($T_\mathrm{ex}$, $N$)$_\mathrm{CO}$ = (1000 K, $2\times 10^{20}$\,cm$^{-2}$)
for 2MASS J055919$-$1404, and
($T_\mathrm{ex}$, $N$)$_\mathrm{CO_2}$ = (800 K, $2.5\times 10^{17}$\,cm$^{-2}$) and
($T_\mathrm{ex}$, $N$)$_\mathrm{CO}$ = (1000 K, $1.5\times 10^{19}$\,cm$^{-2}$)
for SDSS J083008$+$4828, respectively. The results for 2MASS J055919$-$1404
are presented in Figure~\ref{fig:extfit} as an example.
Considering that CO molecules with higher excitation temperature may
located in a slightly interior portion of the photosphere, the ratio of the
column densities 
$N_\mathrm{CO_2} / N_\mathrm{CO} = 2 \times 10^{-2} \sim 3\times 10^{-3}$ 
indicates lower limits of the abundance ratio of the two molecules.
It is found that CO$_2$ is obviously overabundant compared to CO
than predicted by the UCM (Figures~\ref{fig:nlte0559} \& \ref{fig:nlte0830}).
This implies that non-equilibrium processes other than vertical mixing may 
play a role in determining the molecular abundances especially for CO$_2$.

Another possibility is the presence of extended atmosphere beyond 
the hydrostatic photospheres, similarly in the case of red-giants.
It is interesting that \textit{Spitzer}/IRS observation did
not detect any clear CO$_2$ bands in the 15~$\mu$m region
in any brown dwarfs \citep{Roellig04, Cushing06, Saumon06, Mainzer07}.
If the CO$_2$ molecular layer is extended, emission beyond the
photosphere could compensate for the absorption and the observed band
strength would be weakened. This effect is usually more effective
at longer wavelengths. Of course, there is a more conventional explanation
that the heavy opacity of H$_2$O
veils the wavelength region completely. The H$_2$O opacity
reaches a minimum around 4~$\mu$m and makes the 4.2~$\mu$m band visible. 
If this is the case, the CO$_2$ molecules would be located 
in the same layer, or in layers internal to the H$_2$O. 
This may be a constraint on constructing chemical models 
of the brown dwarf atmospheres. 
High dispersion spectroscopy in the 4 \& 15 $\mu$m
region from space is an ideal tool for investigating this problem.

Although non-equilibrium conditions for the CO/CH$_4$ and NH$_3$/N$_2$ ratios
appeared to be explained by the vertical mixing model at least in late-T dwarfs, 
the unexpected detection of CO$_2$ by \textit{AKARI} casts doubt as to
whether the vertical mixing model alone could explain the carbon chemistry even in
late-T dwarfs. In fact, there is no reason why we expect excess
CO$_2$ by the vertical mixing model (Section~\ref{sec:nonlte_midT}) and 
there should be
some other mechanism to produce excess CO$_2$.
A yet unknown mechanism may affect not only
CO$_2$, but also CO as well. In fact, such a new mechanism
could also be what we need to explain the enhancement of CO
in late-L to mid-T dwarfs for which the vertical mixing model failed. 

We continue the \textit{AKARI} brown dwarf programme with the near-infrared
channel of the IRC in the ``post-Helium'' phase (Phase 3).
More targets are being observed to cover the spectral
ranges from M, L, to T. With the enhanced data set we hope to better
understand the some of the current mysteries of ultracool dwarfs.

\acknowledgments
We thank an anonymous referee for critical reading of the manuscript 
and constructive comments especially that helped to revise 
the analysis of Section~\ref{sec:obsvsucm}.
We are grateful to Dr. Gandhi Poshak for his careful checking of 
the manuscript and many suggestions to improve the text.
Dr. Tadashi Nakajima is appreciated for helpful discussions 
at an initial phase in making this observing programme.
Satoko Sorahana gave us useful comments.
This research is based on observations with \textit{AKARI}, a JAXA project
with the participation of ESA.
Issei Yamamura acknowledges JSPS/KAKENHI(C) No.22540260.
Takashi Tsuji thanks to the support by JSPS/KAKENHI(C) No.17540213.
This research has made use of the SIMBAD database,
operated at CDS, Strasbourg, France.



{\it Facilities:} \facility{AKARI (IRC)}.





\end{document}